\documentclass{aastex6}
\usepackage{natbib}
\usepackage{graphicx}

\usepackage[caption=false]{subfig}
\begin{document}

\title{Changing-Look Quasar Candidates: First Results from Follow-up
  Spectroscopy of Highly Optically Variable Quasars}

\author{Chelsea L. MacLeod\altaffilmark{1}}
\author{Paul J. Green\altaffilmark{1}}
\author{Scott F. Anderson\altaffilmark{2}}
\author{Alastair Bruce\altaffilmark{3}}
\author{Michael Eracleous\altaffilmark{4}}
\author{Matthew Graham\altaffilmark{5}}
\author{David Homan\altaffilmark{3}}
\author{Andy Lawrence\altaffilmark{3}}
\author{Amy LeBleu\altaffilmark{6}}
\author{Nicholas P. Ross\altaffilmark{3}}
\author{John J. Ruan\altaffilmark{7}}
\author{Jessie Runnoe\altaffilmark{8}}
\author{Daniel Stern\altaffilmark{9}}
\author{William Burgett\altaffilmark{10}}
\author{Kenneth C. Chambers\altaffilmark{11}}
\author{Nick Kaiser\altaffilmark{12}}
\author{Eugene  Magnier\altaffilmark{11}}
\author{Nigel Metcalfe\altaffilmark{13}}

\altaffiltext{1}{Harvard Smithsonian Center for Astrophysics, 60 Garden St, Cambridge, MA 02138, USA}
\altaffiltext{2}{Department of Astronomy, University of Washington,
  Box 351580, Seattle, WA 98195, USA}
\altaffiltext{3}{Institute for Astronomy, Royal Observatory, Blackford
  Hill, Edinburgh, U.K.} 
\altaffiltext{4}{Department of Astronomy \& Astrophysics and Institute for Gravitation and the Cosmos, 525 Davey
  Laboratory, The Pennsylvania State University, University Park, PA
  16802, USA} 
\altaffiltext{5}{Cahill Center for Astronomy and Astrophysics, California Institute of
Technology, 1216 E. California Blvd., Pasadena, CA 91125, USA}
\altaffiltext{6}{Department of Physics, University of Central Florida,
4111 Libra Drive, Orlando, FL 32816, USA}
\altaffiltext{7}{Department of Physics, McGill University, 3600
  University Street, Montreal, QC, CAN H3A 2T8}
\altaffiltext{8}{Department of Astronomy, University of Michigan, 1085
  S. University Avenue, Ann Arbor, MI 48109, USA}
\altaffiltext{9}{Jet Propulsion Laboratory, California Institute of Technology, 4800
Oak Grove Drive, Mail Stop 169-221, Pasadena, CA 91109, USA}
\altaffiltext{10}{GMTO Corp., 251 S.\ Lake Ave., Pasadena, CA 91101, USA}
\altaffiltext{11}{Institute for Astronomy, University of Hawaii at Manoa, Honolulu, HI 96822, USA}
\altaffiltext{12}{Ecole Normale Supérieure ENS, 24, rue Lhomond, Paris, 75005, France}
\altaffiltext{13}{Department of Physics, University of Durham Science Laboratories, South Road Durham DH1 3LE, UK}

\email{cmacleod@cfa.harvard.edu}

\begin{abstract}
Active galactic nuclei (AGN) that show strong rest-frame optical/UV
variability in their blue continuum and broad line emission are
classified as changing-look AGN, or at higher luminosities,
changing-look quasars (CLQs).  These surprisingly large and sometimes
rapid transitions  challenge accepted models of quasar physics and
duty cycles, offer several new avenues for study of quasar host
galaxies, and open a wider interpretation of the cause of differences
between broad and narrow line AGN.   To better characterize extreme
quasar variability,  we present follow-up spectroscopy as part of a
comprehensive search for CLQs across the full SDSS footprint using
spectroscopically confirmed quasars from the SDSS DR7 catalog. Our
primary selection requires large-amplitude ($|\Delta g|>1$~mag,
$|\Delta r|>0.5$~mag) variability over any of the available time
baselines probed by the SDSS and Pan-STARRS~1 surveys. We employ
photometry from the Catalina Sky Survey to verify variability
behavior in CLQ candidates where available, and confirm CLQs using
optical spectroscopy from the William Herschel, MMT, Magellan, and
Palomar telescopes. For our adopted S/N threshold on variability of
broad H$\beta$ emission, we find 17 new CLQs, yielding a confirmation
rate of $\gtrsim$20$\%$. These candidates  are at lower Eddington
ratio relative to the overall quasar population which supports a
disk-wind model for the broad line region. Based on our sample, the
CLQ fraction increases from 10\% to roughly half as the continuum flux
ratio between repeat spectra at 3420\AA\
increases from 1.5 to 6. We release a catalog of over 200 highly
variable candidates to facilitate  future CLQ searches. 
\end{abstract}


\section{Introduction}

Quasars are known to vary  on the 10-20\% level over time
($\sim$months -- years) and across many wavelengths.  
Their continuum variability is intrinsic to the central
engine, and it causes a response in the broad emission line  (BEL)
flux that is usually lagged and has a smaller amplitude \citep{pet04,ben09}.
 Extreme  variations in optical continuum/BEL flux, 
besides enabling robust studies of the response  between various
emission components, have been caught in recent surveys of the sky 
\citep[e.g.,][]{ste17,ass18} and can reveal rare, interesting physics.
There are
various scenarios for the underlying accretion disk structure that
could explain strong rapid variability in
AGN. These include magnetically elevated (or 
``thick'') disks \citep{dex18},  accretion state transitions
\citep{nod18}, instabilities arising from magnetic torque near
the inner stable circular orbit of the accretion disk \citep{ross18},
 and misaligned disks \cite{nix12}.   

Besides variability, quasar spectra yield information on the structure of the emission regions.
The Balmer lines are thought to provide the most direct measurement
of the amount of ionizing flux relative to the other BELs, and the
H$\beta$ BEL is classically used, along with H$\alpha$, to define
AGN types \citep{ost81}. The profiles of higher ionization lines
seen in the UV (\ion{C}{4}$\lambda\lambda$1548,1550,
\ion{C}{3}]$\lambda$1909) often reveal powerful outflows \citep{ric11}. 
The \ion{Mg}{2}$\lambda$2800 line is generally  less responsive to changes in
ionizing flux than the Balmer lines, presumably due to a larger
average formation radius and/or
because it is intrinsically less responsive \citep{kor04}, but in
recent analysis the line has been observed to 
reverberate among SDSS-RM quasars similarly to H$\beta$ \citep{she16}.  
In the disk-wind model for the BLR described by \citet{eli14}, the dividing line
between a Type 1 and ``true'' (unobscured) narrow-line Type 2 AGN is
a critical value of a parameter:  $L / M_{BH}^{2/3}$, similar to the Eddington
ratio. Conversely, obscured Type 2 quasars can mostly be explained by
varying orientation, as can obscured Type 1 AGN \citep{liut18}.

A small fraction of AGN show large changes in Balmer BELs and are
called ``changing-look'' AGN (CLAGN) if over time they significantly lose or gain
BEL flux 
\citep{kha71,toh76,pen84,coh86,goo89,bis99,sto93,era01,sha14,den14,li15,runco16,hus16}. These
optical CLAGN are different from the classical CLAGN discovered in
the X-rays \citep{ris09}, which are generally associated with large
changes in the X-ray
absorption column.  
Changing-look quasars (CLQs) have recently been defined as  AGN with a
bolometric luminosity $L_{bol}>10^{44}$~erg s$^{-1}$ that exhibit a
strong change in Balmer BELs, usually changing between Types 1 and 1.9
in the optical, and have been recently discovered in surveys
featuring large sample size and long time baselines
\citep{lam15,run15,rua15,mac15,gez17,yang17,ste18}. 
Presumably, CLQs are due
to the same phenomena as optical CLAGN but at higher luminosity and
redshift.  In the handful of cases where there exists X-ray
coverage before and after the event, there are no signs of associated
changes in absorption 
\citep[e.g.,][]{lam15},   
therefore suggesting that strong intrinsic changes in the central ionizing flux
are responsible.

In \citet{mac15}, we performed a systematic search for CLQs through
archival spectra in the Sloan Digital Sky Survey (SDSS) among quasars
with repeat spectra.  This method takes advantage of full-sky
photometric surveys to select a sample of strongly variable
quasars. Now, we use follow-up spectroscopy to uncover additional CLQs.
Recently, \cite{rum17} showed that extremely variable quasars (EVQs,
defined as having $\Delta
g>1$~mag over any baseline) account for 30-50\% of quasars, and they are systematically at
lower Eddington ratio than their less variable counterparts. 
The authors only implied that by extension, CLQs follow this trend.
We are now able to specifically  confirm this trend with our follow-up
spectroscopy. 
We also determine
what fraction of highly variable quasars show strong BEL changes as
a function of continuum flux change using spectral decomposition.

We outline the input photometric data and CLQ candidate selection in
\S\,\ref{sec:phot} and the spectroscopic data and analysis in
\S\,\ref{sec:spec}. We
present our results from spectroscopic follow-up of CLQ candidates,
including the sample demographics and CLQ fraction in \S\,\ref{sec:results}.
Since the purpose of this work is to
systematically search for physically interesting or rare events that
might shed light on the structure in and around accretion disks, we
only highlight the most significant and dramatic cases. However, we
provide the full candidate sample since it may be used for studying
the highly variable tail of quasar variability. 
The results are summarized and discussed in \S\,\ref{sec:disc}.

\section{Photometry and Analysis}\label{sec:phot}
We use imaging data from the SDSS and Pan-STARRS~1
\citep[PS1;][]{kai02} to select strongly variable quasars for
spectroscopic follow-up. In this section we describe the input data 
and our sample selection (see Table~\ref{tab:filter} for a summary).

\subsection{Optical Photometry}
\subsubsection{SDSS}
    The SDSS \citep{York00} uses the imaging data gathered by a dedicated
    2.5m wide-field telescope \citep{Gunn06}, which collected light from a
    camera with 30 2k$\times$2k CCDs \citep{Gunn98} over five broad bands
    - {\it ugriz} \citep{Fukugita96} - to image 14,555 unique
    deg$^{2}$ of the sky. This area includes \hbox{7,500 deg$^{2}$} in the
    North Galactic Cap (NGC) and \hbox{3,100} deg$^{2}$ in the South
    Galactic Cap (SGC).
SDSS started its imaging campaign in 2000 and concluded in 2007, having covered
11,663 deg$^2$. These data are part of the SDSS-I/II survey and are 
described in \citet{DR7} and references therein. SDSS-III added another
$\sim$3000 deg$^2$ of new imaging area in 2008.

    The imaging data are taken on dark photometric
    nights of good seeing \citep{Hogg01}, calibrated
    photometrically \citep{Smith02,ive04, Tucker06, Padmanabhan08a},
    and astrometrically \citep{Pier03} before object parameters are
    measured \citep{Lupton01, Stoughton02}.
    The Eighth Data Release \citep[DR8;][]{DR8} provides updated photometric calibrations.

    The Stripe 82 region of SDSS (S82; 22h 24m $<$ R.A.\ $<$ 04h 08m
    and $|$Dec.$|<1.27$~deg) covers $\sim$300~deg$^2$ and has been observed
    $\sim$60 times on average to search for transient, variable, and moving objects \citep{DR7}.  These
    multi-epoch data have time scales ranging from 3 hours to 8 years and
    provide well-sampled 5-band light curves for an unprecedented number
    of quasars. 

     In our variability analysis and determination of the source
    morphology\footnote{If any epoch has a point-source morphology in SDSS,
     the object is considered a point-source.}, we include all DR10
 primary and secondary photometry as well as 
    observations for point sources in Stripe 82 taken in nonphotometric
     conditions and recalibrated using the improved method of
     \citet{ive04}.

\subsubsection{Pan-STARRS 1 $3\pi$}
  Our analysis includes imaging from the PS1 3$\pi$ survey,
        in particular the Processing Version 2 catalog
        available in a local Desktop Virtual Observatory (DVO) database (released January 2015).
        The overall survey  is described in
        \citet{cha16}; see \citet{mag16,mag16b,mag16c,wat16,fle16} for a description of the data analysis and products. 
        PS1 observations were made with a 1.8m telescope equipped with a 
        1.4-gigapixel camera.  Over the course of 4 years of the 3$\pi$ survey, up to four
        exposures per year in 5 bands, $g_{\rm P1}, r_{\rm P1}, i_{\rm P1},
        z_{\rm P1}, y_{\rm P1}$ have been taken across the full $\delta >
        -30^{\circ}$ sky.  
        Each nightly observation consists of a pair of exposures 15~min apart 
        to search for moving objects. For each exposure, the PS1 3$\pi$ 
        survey has a typical 5$\sigma$ depth of 22.0 in the $g$-band \citep{ins13}.
 Pan-STARRS imaging
commenced in 2009 and continued through to 2013.  Hence, the addition
of the PS1 photometry to the SDSS photometry increases the baseline of
observations from $\approx$8 to $\approx$14 years with typically
several epochs over the overlapping area between the SDSS and ($\sim$30,000
deg$^2$) PS1 footprints.

\subsubsection{Catalina Sky Survey}

 The Catalina Real-time Transient Survey (CRTS) is an open band large-scale survey with dense
monitoring by multiple telescopes of sources brighter than $i\lesssim 19.5$. 
See \citet{dra09} for details.
The CRTS magnitudes are calibrated to a $V$-band zero-point.
Where available, CRTS light curves add the most recent monitoring data
(up to 2016), and the largest number of imaging epochs (typically
$\sim$300 over 10~yr). The data displayed are averaged in 10-day segments.

\subsection{Radio Detections}
To exclude potential jet-related variability from our sample of light
curves, we match to the unified radio catalog of \cite{Kimball_Ivezic14} using a 30\arcsec matching radius.  This
includes data from the FIRST, NVSS, WENSS, GB6 radio surveys, as well
as VLA Low-frequency Sky Survey revised edition (VLSSr).

\subsection{CLQ Candidate Selection} \label{sec:select}

We use optical photometric variability from SDSS and PS1 to select
quasars for follow-up optical spectroscopy. This is a similar
selection as in \cite{mac15}, which was based on $ \Delta g > 1$ mag
changes among SDSS/PS1 photometry and repeat spectra in SDSS and the
SDSS-III's Baryon Oscillation Spectroscopic Survey
\citep[BOSS;][]{Dawson13}. To extend our previous work, we follow up
highly variable quasars lacking recent spectra in BOSS.  For selecting
targets, we limit the sample to redshifts  $z<0.83$ so that the
H$\beta$ line is within the optical wavelength range.

In particular, CLQ candidates are selected by requiring:  
{\it (i)} a magnitude change of $\Delta g > 1$ mag and $\Delta r>0.5$
mag among SDSS and PS1 measurements with errors $< 0.15$~mag, 
{\it (ii)} no radio detection to exclude jet-related variability,
{\it (iii)} a redshift $z<0.83$, 
{\it (iv)} no BOSS spectrum, and
{\it (v)} a current $g$ magnitude at least 1~mag dimmer or brighter
compared to the SDSS spectral epoch. For {\it (v)}, the
photometric data point nearest in time to the earliest SDSS spectrum
is compared to the most recent PS1 data point. 
After visual vetting of the combined SDSS/PS1/CRTS light curves and
spectra, we arrive at 262
CLQ candidates, listed in Table~\ref{tab:clqcans}. 
Table~\ref{tab:filter} summarizes each stage of filtering.
We prioritize these candidates
based on a number of factors to most efficiently use telescope
time. These include:
 their morphology to minimize overlap with the sample from the Time Domain
Spectroscopy Survey (TDSS) ``HYPQSO'' program, which is obtaining 
few-epoch-spectroscopy for hypervariable quasars \citep{mor15,mac18}, but only those with
point-source morphology in SDSS; X-ray detections from the
second release of the {\it Chandra} Source Catalog \citep[CSC;][]{eva18} and/or the 
{\it XMM-Newton} 3XMM-DR5 \citep{ros16} catalog, since they 
afford the opportunity for follow-up with X-ray observatories; 
the S/N ratio in the earlier spectrum;  and the most recent 
photometry from CRTS.  
 While there is no straightforward algorithm to select the 130
observed targets out of the 262 candidates, Table~\ref{tab:filter} gives the
breakdown into three main criteria for  prioritization in the last
three columns, for each stage of filtering.

\begin{table}
  \begin{center}
    \begin{tabular}{lrrrr}
      \hline
      \hline
      Selection     & Total \#  & $i_{\rm SDSS}<19.5$/ &Extended&  {\it Chandra}/\\
           &   & has CRTS  & morphology&{\it XMM} \\
      \hline 
      SDSS Quasars in DR7Q                                           & 105783 &69420&4392&--   \\
      Lacking BOSS spectra                                           & 79838  &53594&4142&--   \\
      EVQs: $|\Delta g|>1$~mag, $|\Delta r|>0.5$~mag                 &        &&&              \\
      \quad ($\sigma<0.15$~mag), $z<0.83$                            & 1727   &1081 &416 &--   \\
      Lacking radio detection                                        & 1403   &858  &332 &103  \\
      $|\Delta g|>1$~mag as of 2013                                  & 262    &203  &86  &20   \\
      Observed spectroscopically {\it (MMT: 64\%, }                  & 130    &105  &67  &17   \\
      \quad  {\it Mag.: 15\%, WHT: 15\%, Pal.: 6\%) }                &        &&&              \\
      CLQs: H$\beta$ (dis)appearance at $N_{\sigma}({\rm H}\beta)>3$  &  17    &15   &10  &0    \\
      \hline
      \hline
    \end{tabular}
    \caption{Selection of spectroscopically variable quasars. Each
     step includes the criteria listed on the previous rows.  The
       rightmost three columns describe prioritized subsets that do not
       together comprise the full sample in the leftmost column. Our X-ray
       catalog is limited to radio-quiet objects.}
    \label{tab:filter}
  \end{center}
\end{table}

\section{Spectroscopy and Analysis}\label{sec:spec}

Starting with a spectroscopic quasar catalog from SDSS, 
we target highly variable quasars for new optical spectroscopy.
We describe the SDSS and follow-up spectroscopic data sets in 
\S\,\ref{sec:specdata}, and describe our spectral analysis in \S\,\ref{sec:qsfit}.
A complete log of the spectroscopic observations and other relevant
information can be found in Table~\ref{tab:clqcans}. 

\subsection{Spectroscopic Data}
\label{sec:specdata}
\subsubsection{SDSS/BOSS}
The final spectroscopically confirmed quasar catalog from SDSS-I/II,
based on the Seventh Data Release of SDSS \citep[DR7;][]{DR7}, is
presented in \citet{sch10}. This catalog contains 105,783 quasars that
have luminosities larger than $M_i = -22.0$. These quasars form our
parent sample, and are hereafter referred to as the DR7Q catalog. 

As described by \citet{ric02}, the bulk of quasar target candidates in
SDSS I/II were selected 
for spectroscopic observations based on their optical colors and 
magnitudes in the SDSS imaging data or their detection in the FIRST 
radio survey \citep{Becker95}. Low-redshift, $z\lesssim 3$, quasar targets
were selected based on their location in $ugri$-color space and the
quasar candidates passing the $ugri$-color selection are selected to a
flux limit of $i = 19.1$. High-redshift, $z\gtrsim 3$, objects were
selected in $griz$-color space and are targeted to $i = 20.2$.
Furthermore, if an unresolved, $i \leq 19.1$ SDSS object is matched to
within 2'' of a source in the FIRST catalog, it is included in the
quasar selection. Additional quasars were also (inhomogeneously) discovered and cataloged 
in SDSS-I/II using X-ray, radio, and/or alternate odd-color information, 
and extending to fiber-magnitudes of about $m<20.5$ \citep[e.g., see][]{anderson03}.

In \citet{mac15}, we looked for significant changes in the BELs of
quasars that had a second epoch of spectroscopy in BOSS, which was
part of the third incarnation of the SDSS
\citep[SDSS-III;][]{Eisenstein11}.   In this work, we target quasars
that lack a BOSS (DR12) spectrum, with two exceptions: two of the CLQs
from \citet{mac15}, J002311.06+003517.5 and J225240.37+010958.7, were targeted for follow-up
spectroscopy here because they  showed a strong dimming in PS1 since
the BOSS spectrum.

\subsubsection{William Herschel Telescope (WHT)}
Relatively bright ($g<20.5$) and highly variable ($|\Delta
g|>1.3$~mag) CLQ candidates were observed on the nights of 2016 February 6-8 and May 30,
31 using the 4.2m William Herschel Telescope (WHT) in La
Palma. Observations were performed using the Intermediate dispersion
Spectrograph and Imaging System (ISIS). The 5300 dichroic was used
along with the R158B and R300B gratings in the red and blue arms,
respectively, along with the GG495 order sorting filter in the red
arm. Typically 2$\times$ binning in the spatial direction was used to
improve the signal-to-noise ratio (SNR) along with a narrow CCD window
to reduce memory usage and readout times. This set-up gives a spectral
resolution of $R\sim$1500 at 5200\,\AA\, in the blue and $R\sim$1000
at 7200\,\AA\, in the red for a slit width of 1\farcs{0} and nominal total
coverage of $\sim3100\textup{--}10600$\,\AA\  (but effectively further limited by the atmosphere).

Typically, calibration images were taken at the start of each night
including bias frames, lamp flats and CuNe/Ar arc lamp
images. Spectroscopic standard stars were observed at $\sim$2h
intervals throughout the night though this cadence was not always
possible. The slit was oriented at the parallactic angle. Exposures
were taken in $1800{\rm s}$ increments and the number of shots on
target was adjusted based on the latest PS1 photometry. WHT data were
reduced using custom \textsc{pyraf} scripts and standard techniques.

For analysis of follow-up spectra from WHT and other telescopes
described below, we correct for telluric absorption where needed by
using a standard star observation at similar airmass and the empirical
method described in \citet{wade88} and \citet{oster90}.

\subsubsection{MMT}
Observations of a fainter set of targets were made with the Blue
Channel Spectrograph on the 6.5m MMT situated on Mount Hopkins,
Arizona. Observations were carried out over several dates in 2016-2018
(see Table~\ref{tab:clqcans} for exact dates).  Here, the
300 $\ell\, {\rm mm}^{-1}$ grating was used with a clear filter and
$2\times$ binning in the spatial direction. The central wavelength
setting ranged from 5835\AA\ to 6335\AA\ (mostly at 5900\AA). To
reduce MMT data, we used the {\tt pydis} software adapted for use with 
Blue Channel data.\footnote{{\tt http://jradavenport.github.io/2015/04/01/spectra.html}} 
We use three exposures to remove cosmic rays by taking the median, and
we compute the error in flux as rms$/\sqrt{N}$. When three exposures were not
available, we use the errors output by  {\tt pydis} and remove cosmic
rays by filtering large deviations from the background regions on the
smoothed two-dimensional spectrum. The Blue Channel spectrograph (on
MMT) achieves good signal at observed-frame wavelengths
$\lambda<7000$\AA. 
  
\subsubsection{Magellan}
To observe most of our targets in the South Galactic Cap (SGC), and
most of the relatively high-redshift targets, we 
used the Magellan Clay 6.5m telescope with the Low Dispersion Survey
Spectrograph 3 (LDSS3)-C spectrograph. Observations were carried out
over the nights 2016 July 26-29. The LDSS3-C instrument consists of a grism situated
behind an aperture plate; we opted to use the VPH-All grism (covering
4250--10000\AA) with the standard $1\farcs0 \times 4\farcm0$ center
long slit mask.

The LDSS3-C spectrograph has better efficiency
toward redder wavelengths ($7000<\lambda<10000$\AA) than the MMT's
Blue Channel, so we preferred it for higher-redshift objects. 
Reductions were carried out both using standard IRAF techniques, and
the {\tt pydis} software adapted for use with LDSS3 data.

\subsubsection{Palomar}
We obtained second-epoch spectra of two quasars with Double
Spectrograph (DBSP) on the Hale 200'' Telescope at Palomar Observatory
on UT 2017 May 30, and second-epoch spectra of an additional six 
quasars using the same instrument on UT 2017 June 26.  All quasars 
were observed with single 900~s exposures, other than J161602.39+482201.2
which was observed with two such exposures.  We obtained all the
spectra through a 1\farcs5 slit aligned at the parallactic angle
using the 600 $\ell\, {\rm mm}^{-1}$ grating on the blue arm of the
spectrograph ($\lambda_{\rm blaze} = 4000$~\AA), the 316 $\ell\,
{\rm mm}^{-1}$ grating on the red arm of the spectrograph ($\lambda_{\rm
blaze} = 7500$~\AA), and the 5500~\AA\ dichroic.  This configuration 
provides moderate resolution spectra across the entire optical 
window.  The data were processed using standard techniques within
IRAF, and flux calibrated using standard stars observed on the same
night.

\subsection{Spectral Decomposition}
\label{sec:qsfit}
We use the QSfit spectral decomposition code from \citet{cal17} to
analyze the SDSS and follow-up spectra. We fit simple power-law
continuum components, iron optical and UV templates,
Balmer continuum, broad emission line, narrow emission line, and
host galaxy components. The \ion{Fe}{2}  templates are based on the narrow-line
Seyfert 1 (NLS1) AGN I Zw 1 \citep{ves01,ver04}, with the UV iron
equivalent width (EW) fixed to avoid a degeneracy with the 
continuum slope. The host galaxy template is a typical AGN host
\citep{sil98,pol07}. We adopt the redshifts and $E(B-V)$
values from the DR7Q catalog. We exclude spectra where the median S/N
is $\lesssim 2$. For an example spectral decomposition near H$\beta$,
see Figure~\ref{fig:qsfiteg}. 

For the H$\beta$ line, we start by fitting a single broad plus a
single narrow component with both FWHM values allowed to vary. The
FWHM of the broad component was  limited to be between
900-15,000~km~s$^{-1}$, as in \citet{cal17}. The narrow lines were
modeled with single components with a FWHM limited to the range 100-2000~km~s$^{-1}$.
For some objects, the broad H$\beta$ component was flagged
as having bad quality (e.g., the FWHM or velocity offset hit a
limit). In these cases, if there was no apparent broad H$\beta$ flux (which
was usually the case), we fixed the FWHM of the broad component to the
full width at 10\% maximum of the broad component from the bright-state
spectrum to find an upper limit on the H$\beta$ flux.\footnote{
    Our results are not affected whether we use the full width 
  at 10\% maximum or the FWHM of the bright-state H$\beta$ flux, for
  the CLQ dim state flux.}

After fitting each spectrum, we scaled the follow-up spectrum by a
single factor so that the modeled sum of narrow
[\ion{O}{3}]$\lambda\lambda$4959,5007\AA\ line flux
(continuum-subtracted) matched that of the early-epoch SDSS spectrum. 
In some cases where the [\ion{O}{3}] scaling yielded unphysical
results, we instead forced the host galaxy component to match that of SDSS.   

Then, using the Monte Carlo resampling technique from  \citet{cal17},
we resampled 100 times, taking the rms of the resulting parameters for
their uncertainties.

\subsection{Definition of CLQs}

\begin{figure}[h!]
\centerline{
\includegraphics[scale=.4]{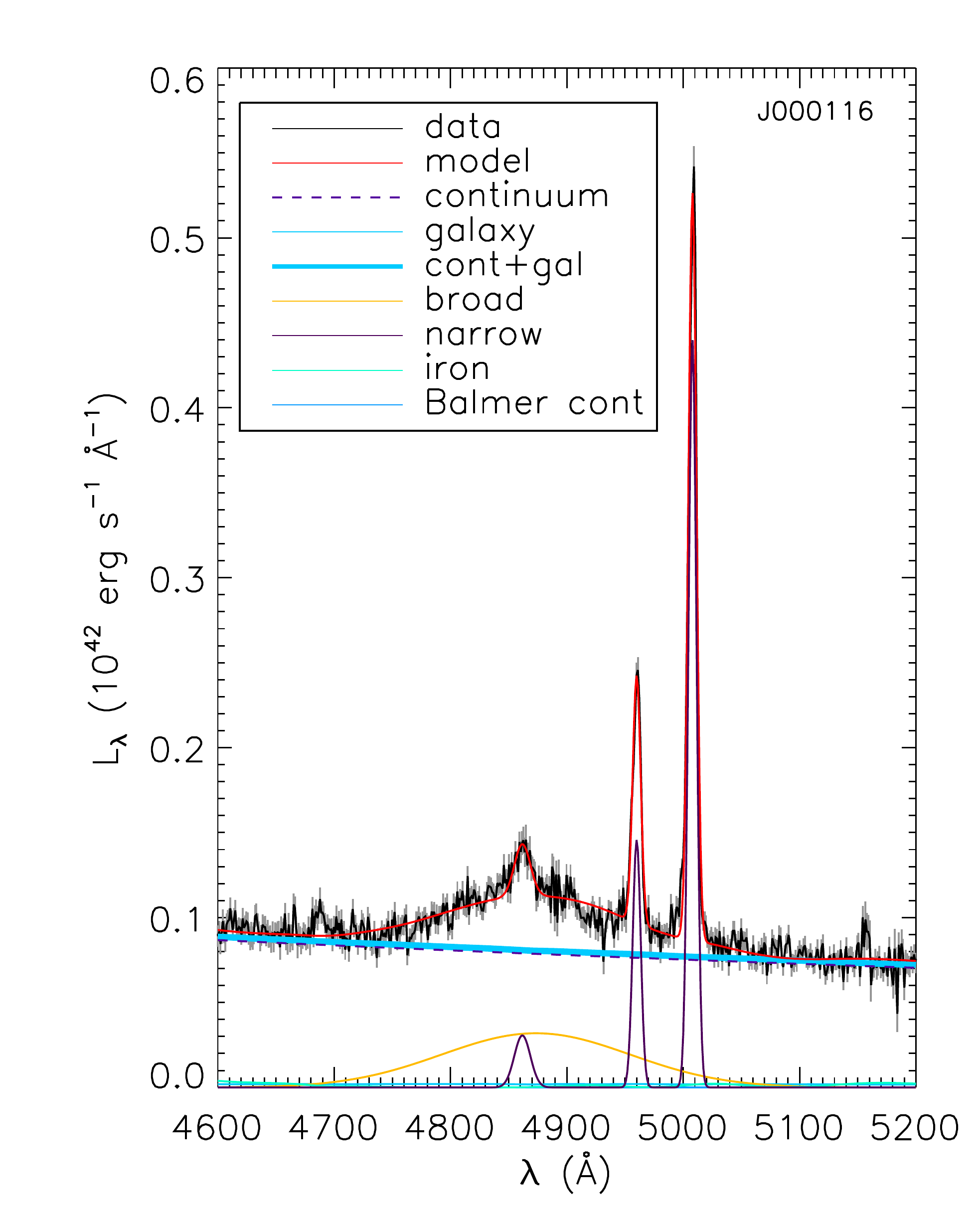}
\includegraphics[scale=.4]{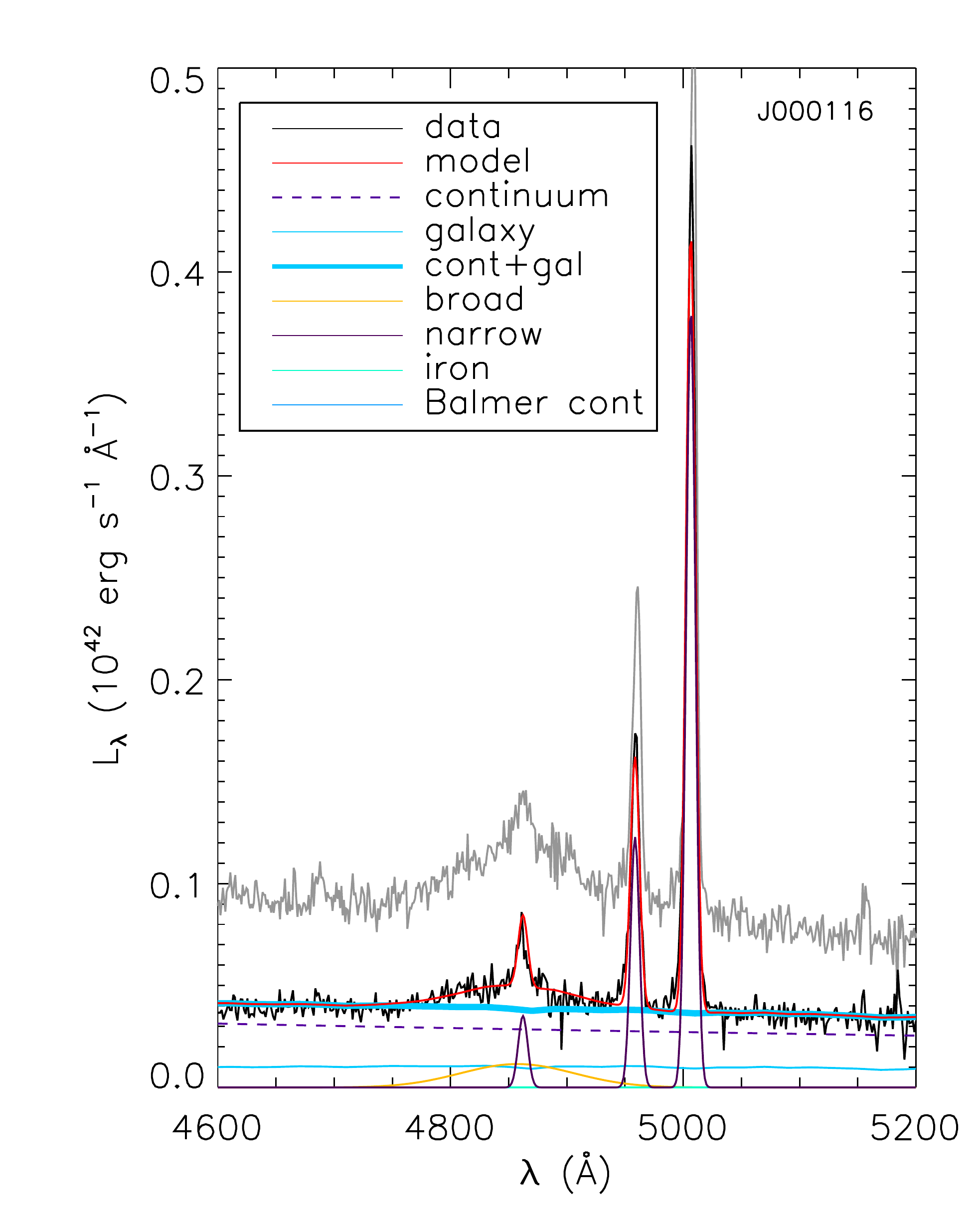}}\centerline{
\includegraphics[scale=.4]{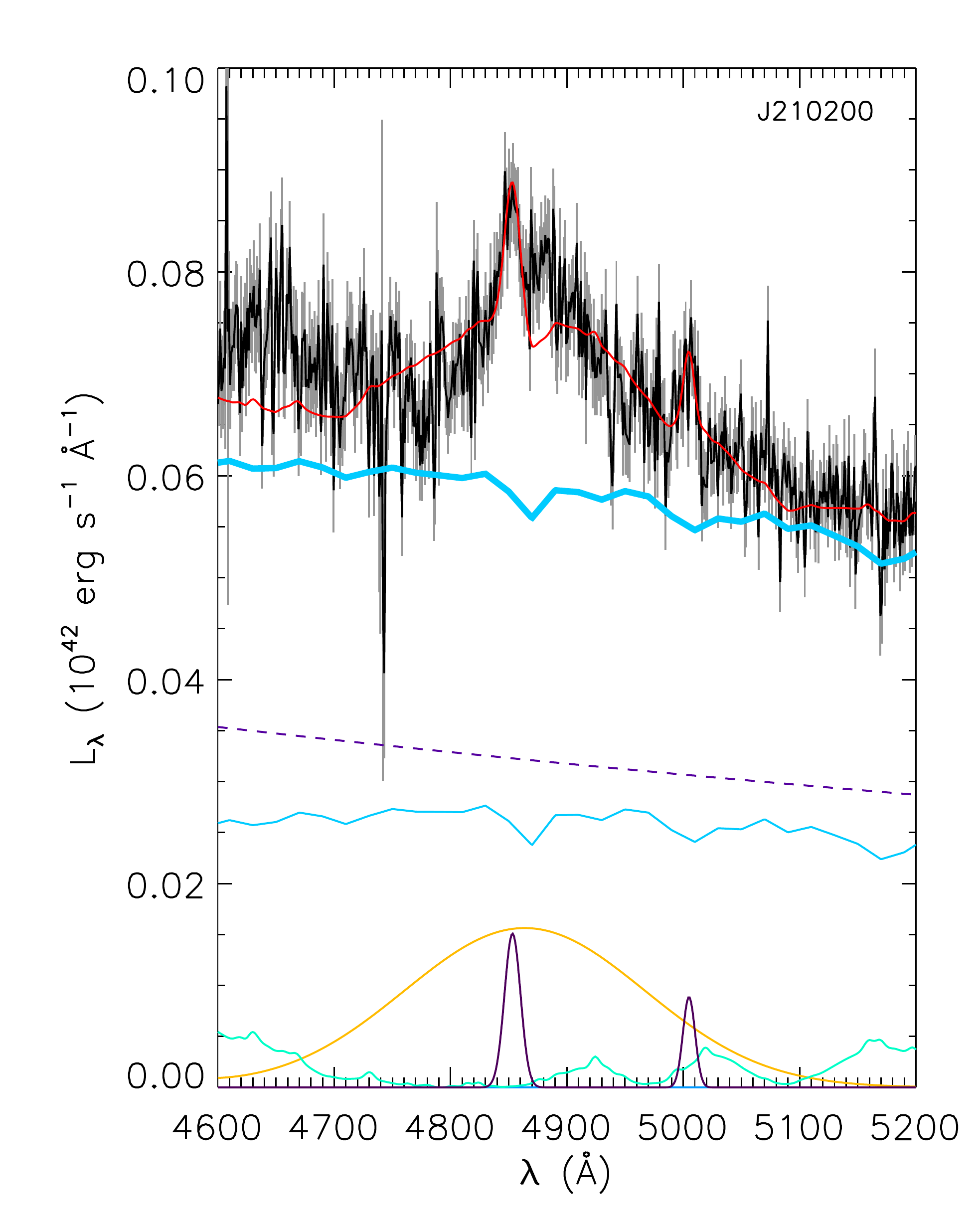}
\includegraphics[scale=.4]{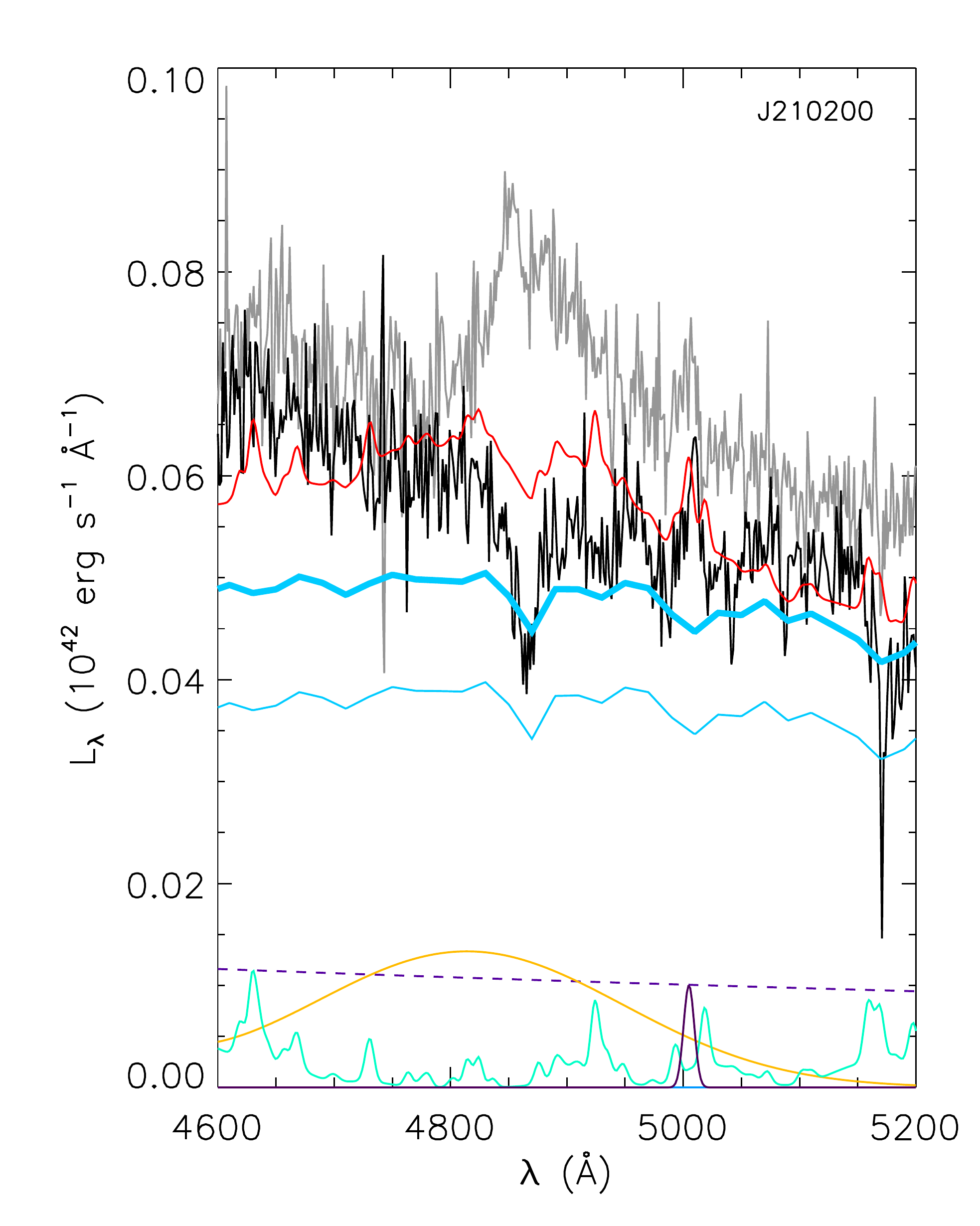}}
\caption{\footnotesize{
\{it Top panels:} example of a spectral decomposition for variable quasar J000116.00+141123.0
showing an intermediate luminosity change ($|\Delta L_{3240}|\simeq
3.5\times 10^{44}$~erg~s$^{-1}$) and
retaining some broad H$\beta$ flux.  This object is not classified as
a CLQ here because there is still broad H$\beta$ visible in the dim
state, and the significance of the BEL change is $N_{\sigma}({\rm
  H}\beta)<3$.
 In the left panel, the SDSS spectrum, best-fit model,
and model components are shown.  In the right panel, the SDSS spectrum
is shown in gray, and the spectral decomposition for the follow-up
spectrum, in black, is shown. The follow-up spectra are scaled so that
the integrated [\ion{O}{3}]$\lambda\lambda$4959,5007\AA\ model narrow
line flux matches that of the SDSS spectrum. 
 {\it Bottom panels:} as in the top panels but  for 
J210200.42+000501.8, a changing-look quasar in a post-starburst galaxy
\citep{cal13}. In this case, the BEL component
fit to the dimmer spectrum was fixed to 10\% the SDSS flux and is
considered only an upper limit.
\label{fig:qsfiteg}}
\end{figure}

The S/N can have a large impact on the visual definition of CLQs.  
If the S/N were worse for certain objects, e.g., J000116.00+141123.0 shown
in Figure~\ref{fig:qsfiteg}, they might be considered CLQs
upon visual inspection since H$\beta$ would be hidden in the noise. 
To limit the number of cases that simply lacked the S/N to rule out a
CLQ nature, we use the significance of the H$\beta$ change to form our CLQ sample.

We calculate the flux deviation,
\begin{equation}
N_{\sigma}(\lambda)=(f_2-f_1)/\sqrt{\sigma_2^2+\sigma_1^2}
\end{equation}
 to determine the significance (in units of
$\sigma$ per spectral element) for the H$\beta$ change. 
First, we scale the spectra as described in \S\,\ref{sec:qsfit} and 
subtract each continuum component to remove the linear trend across
the line.  Then, we re-bin the spectra to a similar resolution in
  rest-frame wavelength of about 2\AA/pix. Finally, we smooth the
  resulting spectra using a running median with a window of 32\AA\
  (rest-frame). The most extreme value of $N_{\sigma}(\lambda)$ in the
wavelength range of H$\beta$ (4750--4940\AA\ rest-frame),  indicative
of the amount of H$\beta$ variability, is then compared to 
$N_{\sigma}(4750{\rm \AA})$  (i.e., the flux deviation just blue-ward of
H$\beta$).   Note that the value $N_{\sigma}(4750{\rm \AA})$ is
  determined from the smoothed array so that we are not susceptible
  to outliers in flux at that wavelength. 
This method is designed to detect H$\beta$ variability relative to the
  4750\AA\ continuum:
if the maximum
$|N_{\sigma}(\lambda)|$ of the line exceeds the flux deviation in the
neighboring continuum  by a certain amount, the object may be
classified as a CLQ. We compute the maximum flux deviation of the line
relative to the continuum $N_{\sigma}(4750-4940{\rm \AA}) -
N_{\sigma}(4750{\rm \AA})$, which we express as $N_{\sigma}({\rm
  H}\beta)$.

We define CLQs as having a visual disappearance (or emergence) of
broad H$\beta$ at a $N_{\sigma}({\rm H}\beta)>3$ level.
In Table~\ref{tab:clqcans}, the visually identified CLQs are indicated
by a ``1'', and $N_{\sigma}({\rm H}\beta)$ is also provided.
At least half of the apparent ``CLQs'' from our visual inspection had
$N_{\sigma}({\rm H}\beta)<3$; 
these objects would benefit from better quality spectra to confirm or
refute the absence of broad BEL components. 

We note that by using a $N_{\sigma}({\rm H}\beta)$ definition, we may
identify as CLQs objects whose broad H$\beta$ does not change by a
large amount. However, our definition is primarily a practical one, as
it uses quantities that we can readily measure. An alternative criterion
such as an absolute or relative change in broad H$\beta$ flux would
lead to a different set of ambiguities. Moreover, so far CLQs have
most often  been identified by visual inspection of the spectra
without a quantitative definition. Therefore, we adopt this
straightforward definition and  explore the consequences.

\section{Results} \label{sec:results}

Using the WHT, MMT, Magellan, and Palomar, follow-up spectra have been obtained
for 130 sources.  In Table~\ref{tab:clqcans} we list all objects
selected as CLQ candidates, and indicate those already observed with follow-up
spectroscopy. Included in the Table are the two $g$-band
photometric measurements leading to the $ \Delta g > 1$ mag selection
($g_1$ and $g_2$ with associated errors $\sigma_1$ and $\sigma_2$),
along with their MJDs (MJD$_1$ and MJD$_2$). We also list the most
recent $g$-band magnitude ($g_{\rm PS1}$) and spectral epochs (SDSS and follow-up).

The repeat spectra for the highest-confidence CLQs with $N_{\sigma}({\rm
  H}\beta)>3$, numbering at 17,  and the corresponding light curves
are shown in Figure~\ref{fig:clq1}. Lower-confidence CLQs with $1<N_{\sigma}({\rm
  H}\beta)<3$ are shown in the Appendix (Figure~\ref{fig:clqA1}). 
The light curves suggest a strong dimming (or brightening) between the
spectral epochs, but in most cases there is not enough photometric
sampling to resolve the timescale for the  change.  The CRTS data 
resolve the time of transition more often than the  SDSS/PS1 light
curves, but due to the unfiltered bandpass, the CRTS light curves are
less sensitive than the $g$-band to blue
continuum variability.
The results from light curve modeling \citep[following][]{koz10,mac10}
and characterization \citep[following][]{kim11}  of the SDSS/PS1 or CRTS
photometry essentially confirm our selection of CLQ candidates: that
they have a large change in flux at some point between the SDSS and
Pan-STARRS data (leading to longer characteristic timescales,
e.g.). Among those CLQ candidates that we followed up
spectroscopically, we notice no overall difference between the light curves of
CLQs versus non-CLQs.

\begin{figure}[h!] 
  \subfloat[][Fig. 2$a$]{\centerline{
      \includegraphics[scale=.4]{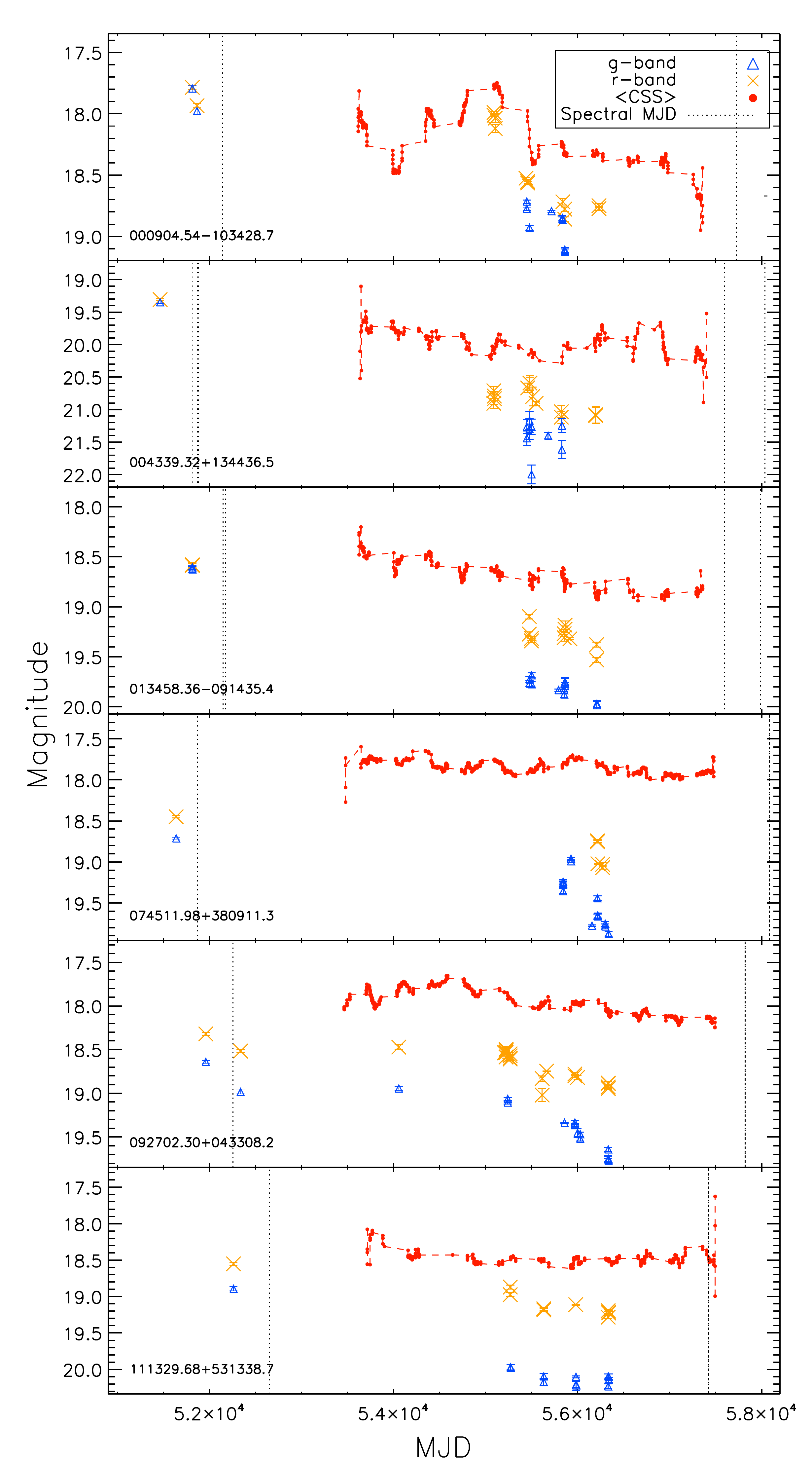}
      \includegraphics[scale=.4]{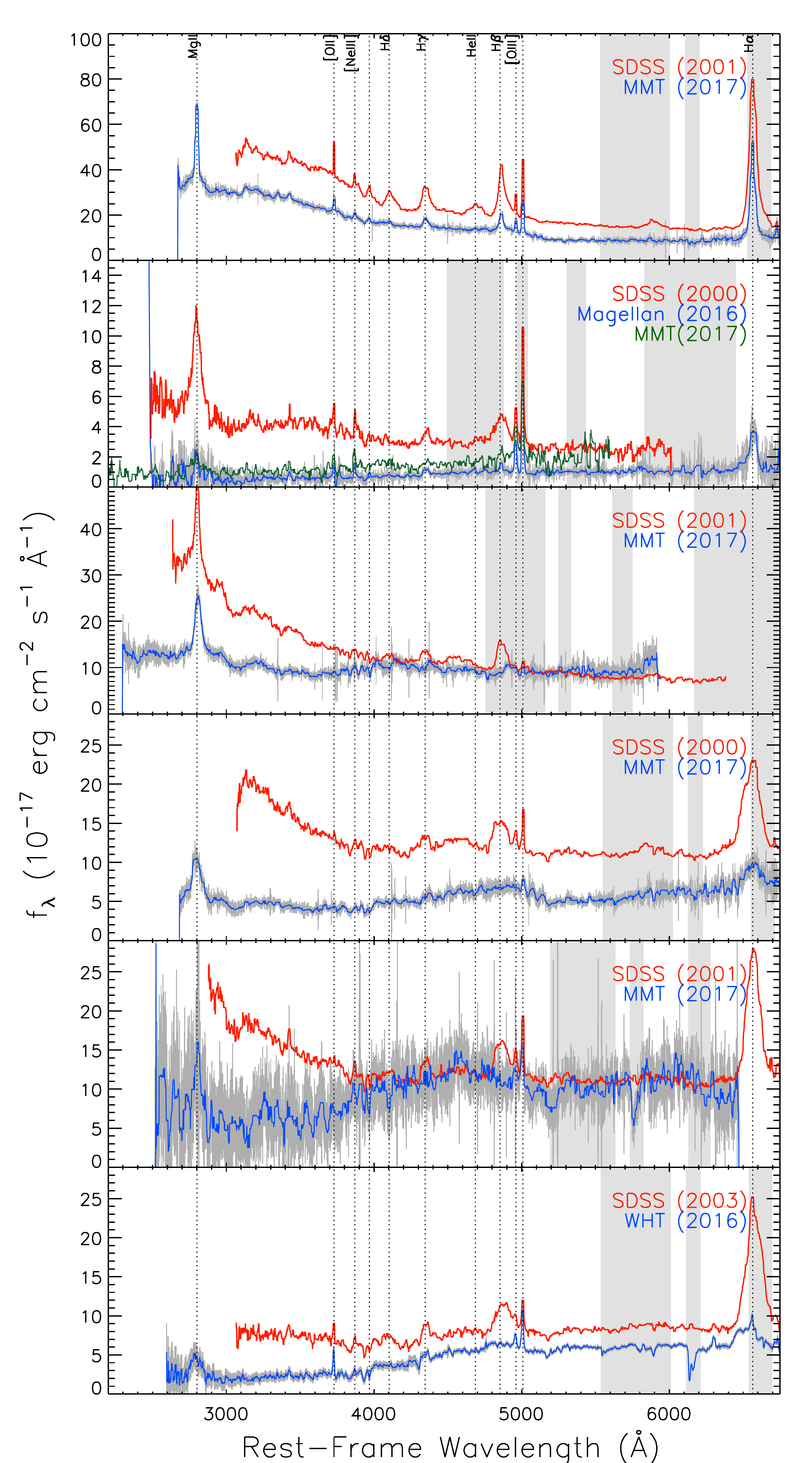}}}
  \caption{\footnotesize{ CLQ light curves and repeat spectra shown in R.A.\ order. 
      \emph{Left:}   SDSS and Pan-STARRS $g$-band ($r$-band) photometry is shown
      as blue triangles (orange crosses). Red data points show  
      archival photometry from CRTS in unfiltered
      light, averaged every 10 days.
      The existing spectroscopic epochs are indicated by the vertical lines. 
      \emph{Right:} Existing spectra for objects in the adjacent left panel (red is
      SDSS;      
      blue or green is our follow-up), corrected for telluric
      absorption where needed. The telluric bands are shown in grey,
      as are the error bars in flux.
      The follow-up spectra plotted in green improve the coverage to
      shorter or longer wavelengths, but are not included in the
      analysis (their errors are not shown).   
    }\label{fig:clq1}}
\end{figure}
\begin{figure}[h!]                 
  \ContinuedFloat
  \subfloat[][Fig. 2$b$. CLQs, continued. ]{\centerline{
      \includegraphics[scale=.4]{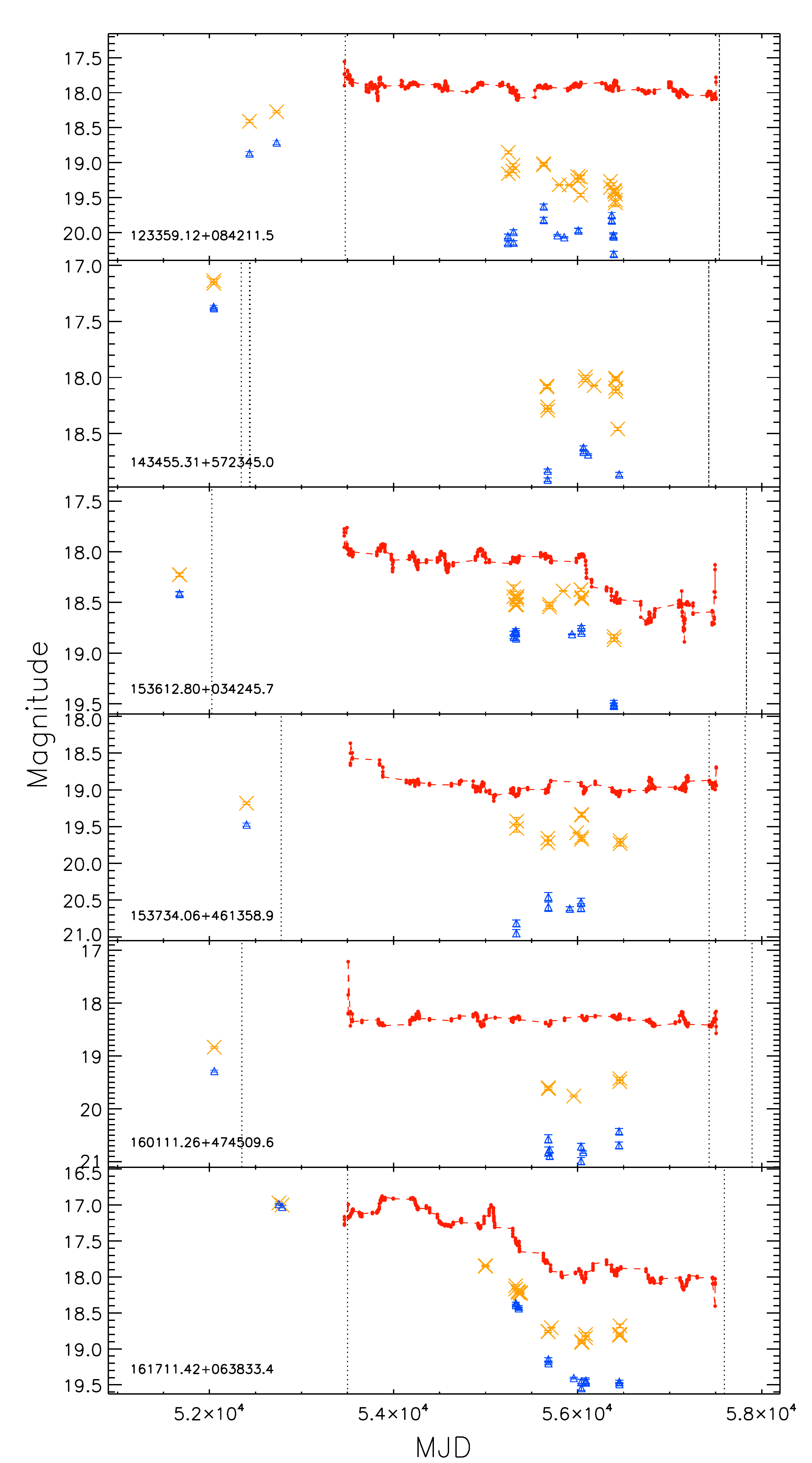}
      \includegraphics[scale=.4]{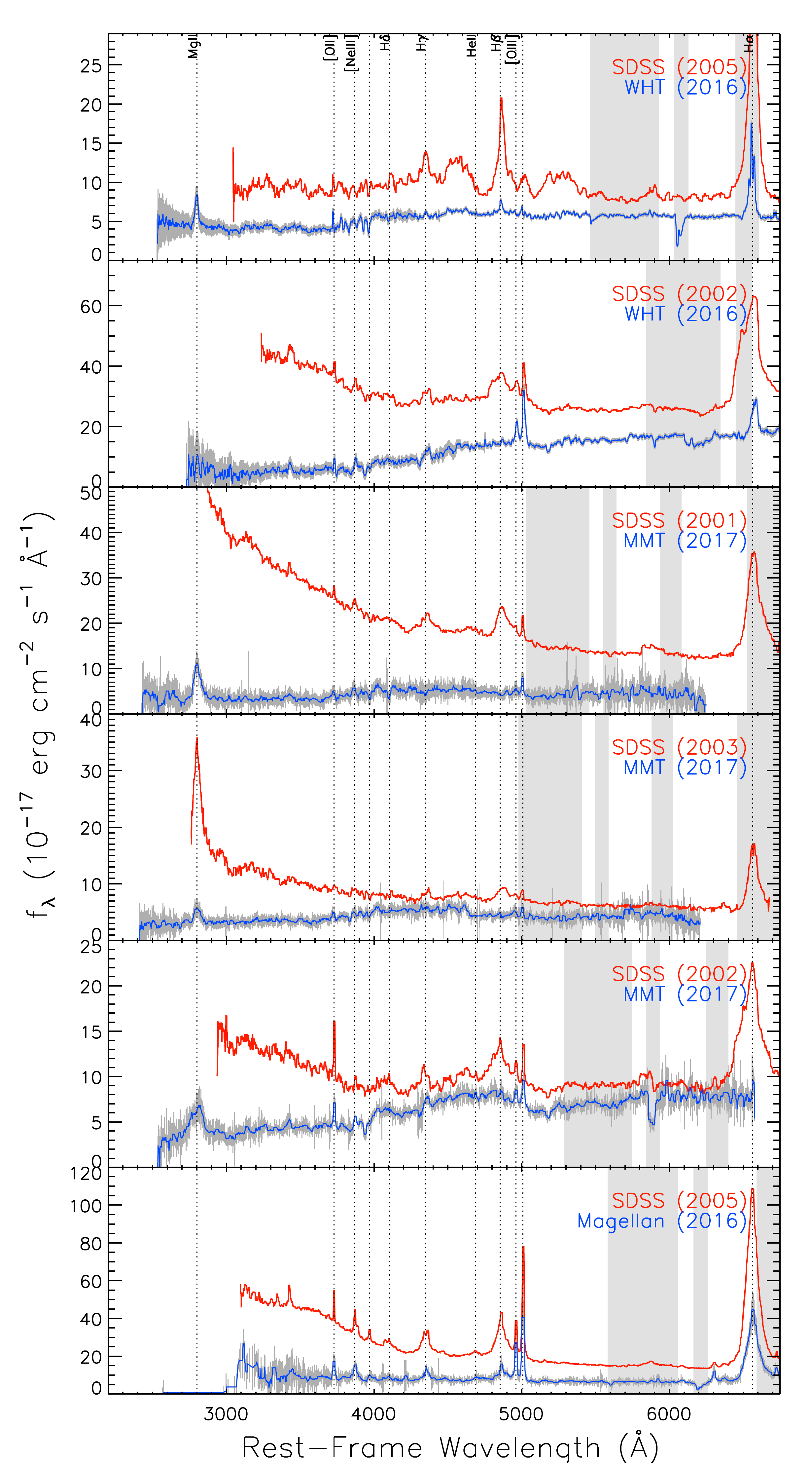}}}
\end{figure}
\begin{figure}[h!]                 
  \ContinuedFloat
  \subfloat[][Fig. 2$c$]{\centerline{
      \includegraphics[scale=.4]{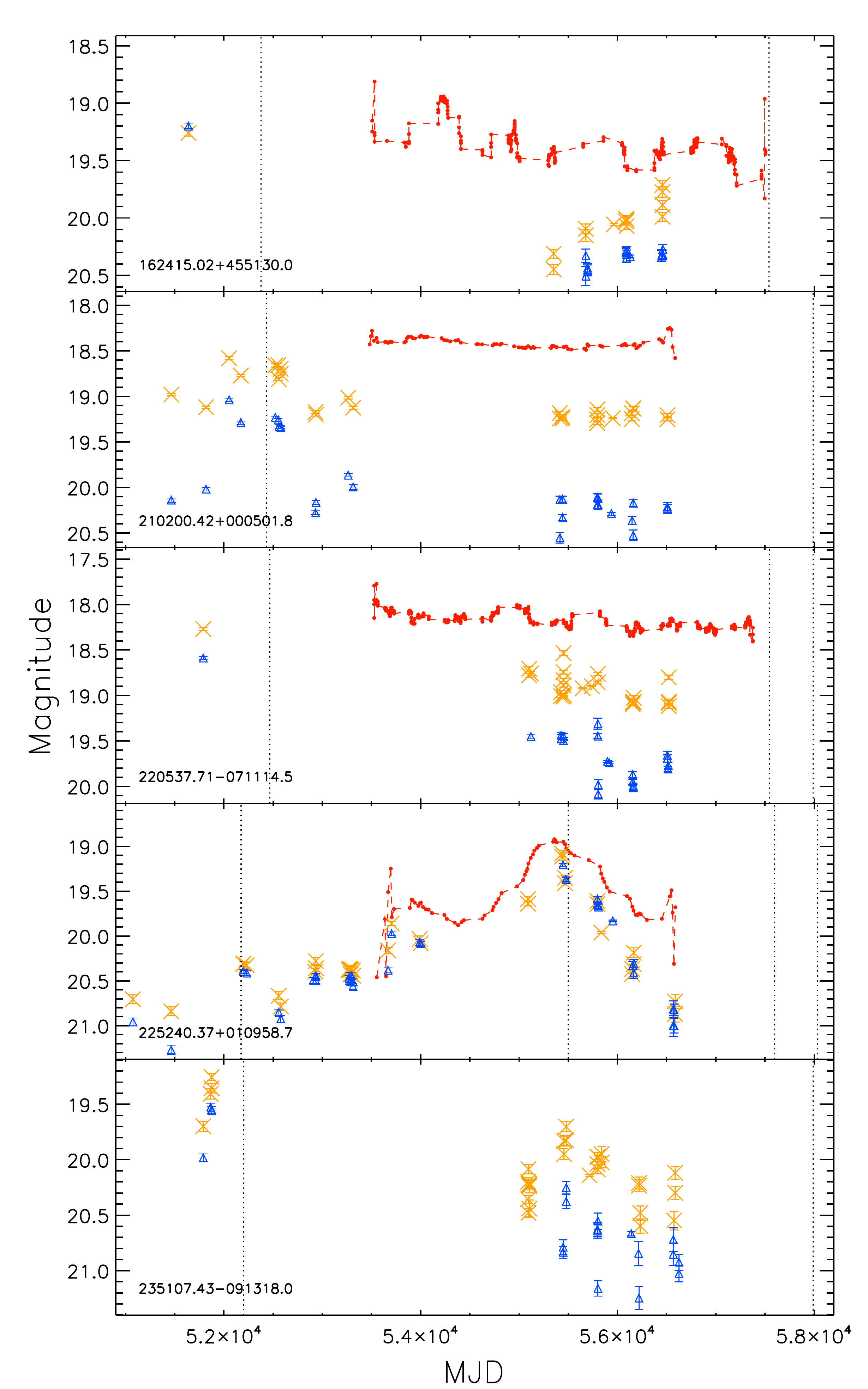}
      \includegraphics[scale=.4]{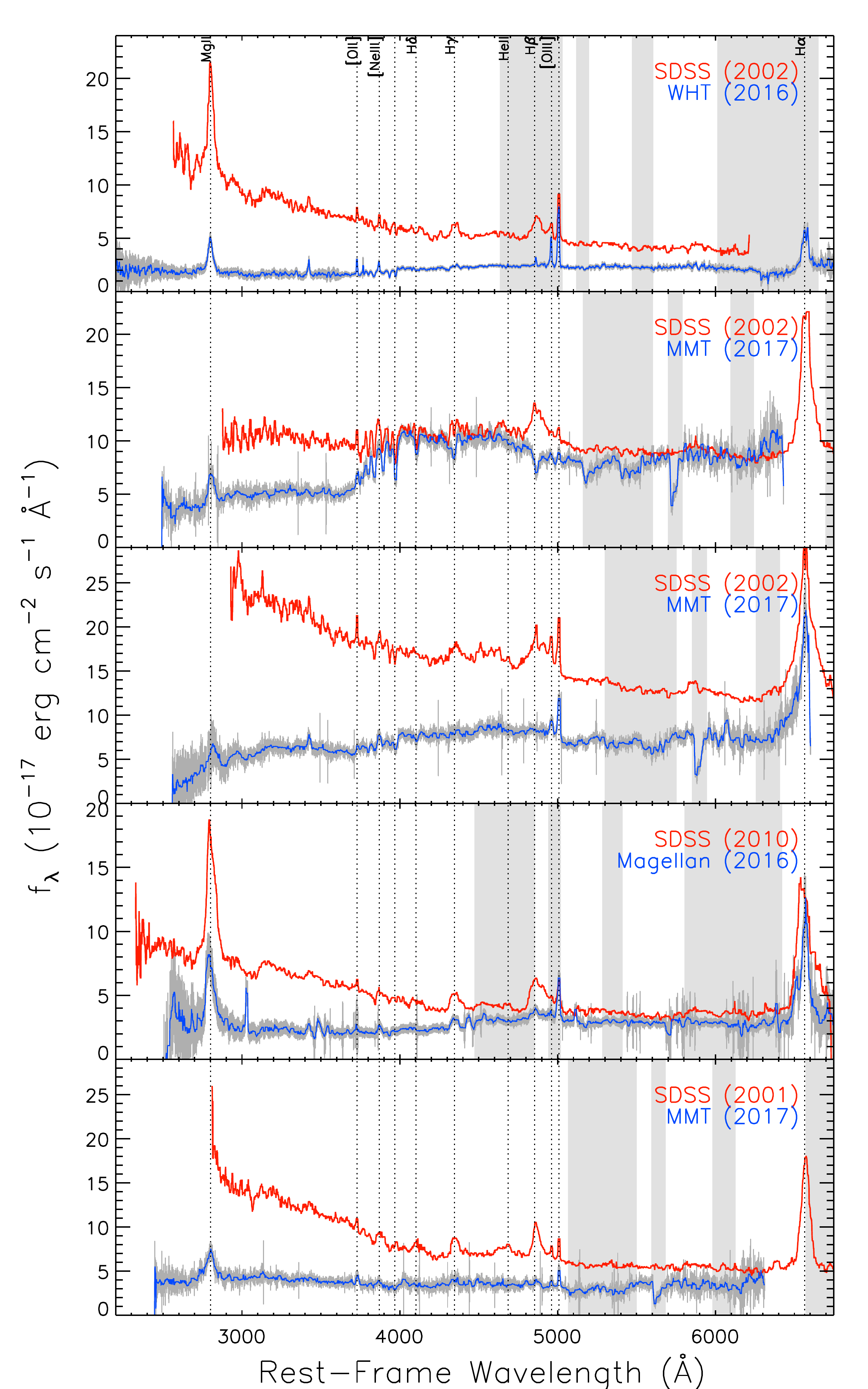}}}
\end{figure}
\setcounter{figure}{2}

\subsection{Distribution of Luminosity Changes and Flux Ratios}

The luminosity change at 3240\AA, calculated from the simple
power-law continuum component from {\tt QSfit} (\S\ref{sec:qsfit}), is
plotted against the luminosity change of H$\beta$ in
Figure~\ref{fig:qsfit}.  
The uncertainties were determined from the rms of model component
fits from 100 Monte Carlo realizations of the data; then the errors
in luminosity from both epochs are added in quadrature to compute the
error in $|\Delta L|$. 
The flux changes are shown in the adjacent panel. There is a positive
correlation between continuum change and BEL change whether measured
in luminosity or in flux.

 The final CLQ sample is shown by red diamonds in Figure~\ref{fig:qsfit}.  
We find that CLQs are interspersed throughout the entire
distribution. There also exist cases where either the broad components
did not vanish completely, despite substantial continuum luminosity
change, or the signal to noise was not deemed sufficient for a CLQ
classification (e.g., see \S\,\ref{sec:nearclqs}).  
When multiple follow-up spectra exist, the spectrum with highest S/N
was chosen for the comparison with SDSS and is listed in
Table~\ref{tab:clqcans}.  The resulting luminosities are given in
Table~\ref{tab:qsfit}. 
 
To determine more clearly how the line varies relative to the continuum, 
we compute the flux ratio between the bright and dim states.
If the flux ratios  are  proportional  to  each  other, that would
mean a constant EW and suggest a constant spectral energy
distribution (SED) during the transition.  
Figure~\ref{fig:qsfitfrac} shows the flux ratio  $\lambda L_{3240,
  high} / (\lambda L_{3240, low})$ against $L_{H\beta, high}/L_{H\beta, low}$. 
We see that strong Balmer line variability is associated with the
largest continuum variations. Given the number of lower limits and
the large scatter, it is not clear whether the relative flux changes are
proportional without a larger sample.
The marginal distributions are shown for both axes so that the
location of CLQs relative to the tail can more easily be seen.
The CLQ fraction is
displayed as a function of  $\lambda L_{3240, high}/(\lambda L_{3240, low})$ in the
right panel of Figure~\ref{fig:qsfitfrac}.

\subsubsection{Highly Variable Quasars and Intermediate Transitions}\label{nearclqs}
Among the objects that do not exhibit strong BEL changes in the
follow-up spectrum, in many cases this is due to a rebrightening
since the last PS1 epoch, as revealed by CRTS photometry.  Roughly
half the observed cases were found to have recent brightening by
several tenths of a magnitude. When excluding the rebrightened
sources, we find a CLQ confirmation rate of 20-50\%  for flux
deviations $N_{\sigma}({\rm H}\beta)>$3-1, respectively, from our
follow-up.  

Among the non-CLQs that \emph{did} show a strong continuum change, 
in some, the BEL flux remained after clearly responding to the change
in the continuum level,
e.g., J000116.00+141123.0 and J012946.71+150457.2
(Figures~\ref{fig:qsfiteg} and~\ref{fig:obj_interest}). 
In \S\,\ref{sec:asym}, we elaborate on the latter object.
See Figure~\ref{fig:nearclqs} for more objects
with significant continuum and BEL variability that still retain a
small portion of the broad H$\beta$  line.
For example, the object J233843.40-105719.5 in the last panel of
Figure~\ref{fig:nearclqs} lost all the 3240\AA\ power-law continuum flux.
This object might be classified as a Type 1.8, since it
still has broad H$\alpha$ and a slight amount of broad H$\beta$ in the
dim state.

In several other candidates,   Balmer BEL flux was retained in the dim
state simply because the \emph{fractional} luminosity change was
relatively small; i.e., they were very luminous before dimming. 
 For this reason, we find that the largest change in H$\beta$ or
  3240\AA\ luminosity is
a non-CLQ (J025619.01$-$004501.3 at $z=0.72$; Figure~\ref{fig:qsfit} left panel). However, this changes 
when instead considering the flux ratio between high and low states (Figure~\ref{fig:qsfitfrac}).
Moreover, by comparing Figures~\ref{fig:qsfit} and
\ref{fig:qsfitfrac}, the CLQ fraction is more sensitive to the
fractional continuum luminosity change rather than the absolute
change.

\subsection{Other Behavior and Objects of Interest}

\subsubsection{Strong \ion{He}{2}$\lambda$4686\AA\ variability}
\label{sec:He2}
Figure~\ref{fig:clq1} (top panel) shows an example of a
CLQ (J000904.54-103428.7) with vanishing BELs, and a particularly strong change in
\ion{He}{2}$\lambda$4686\AA. 
In reverberation mapping studies of NGC 5548 \citep{fau16},
\ion{He}{2}  is seen to respond quickly and strongly to continuum
variations, implying that the size of the \ion{He}{2} emission region
is very close to the ionizing continuum source.
A strong  \ion{He}{2} change is also seen in Mkn 110 \citep{kol02},
and the source J013203 shown in Figure~\ref{fig:obj_interest},
although broad H$\beta$ is still clearly present in the dim spectrum. 

\subsubsection{NLS1 AGN: J123359.12+084211.5}

J123359.12+084211.5 (Figure~\ref{fig:clq1}) shows a remarkable change in
\ion{Fe}{2} emission in a NLS1 AGN. See \cite{bla17} for a similar
spectral change in a NLS1. J123359.12+084211.5 has a relatively small continuum
change and large Eddington ratio compared to the other CLQs, making it
a noticeable outlier.   

\subsubsection{The Post-Starburst Quasar J210200.42+000501.8}
The source J210200.42+000501.8 (Figures~\ref{fig:qsfiteg} and \ref{fig:clq1})
was classified in  \citet{cal13} as a post-starburst 
quasar.  The authors presented a follow-up spectrum taken prior to our
MMT spectrum when the quasar was at an intermediate stage between the
two spectra listed here, based on the light curve. The SDSS spectrum
was obtained during a flaring event. Another dramatic CLQ,  J101152.98+544206.4
\citep{run15}, was found in a post-starburst galaxy, but the
light curve bright phase exceeded a short-lived flare, leading the authors
to conclude that it was not a tidal disruption event or a supernova.

\subsubsection{Flickering CLQs J002311.06+003517.5 and J225240.37+010958.7}\label{sec:flicker}
The objects  J002311.06+003517.5 and J225240.37+010958.7   were both presented as CLQs in
\cite{mac15}. They both showed a significant dimming in PS1 since the
latest BOSS spectra, so we observed them with Magellan
in July 2016 and with MMT in October 2017.
The emerging and disappearing Balmer BELs appear to
be associated with dramatic brightening and dimming in the light curves, and they have
recently turned off again (see Figures~\ref{fig:clq1} and \ref{fig:clqA1}). 
J225240.37+010958.7 has a boxy H$\beta$ profile (see \S\ref{sec:asym}), and exhibits only one dramatic
flare over the course of the light curve, whereas J002311.06+003517.5 shows
persistent large-amplitude variability throughout its light curve.

\subsubsection{Asymmetric broad Balmer line profiles}
\label{sec:asym}
\begin{figure}[h!]
\centerline{
\includegraphics[scale=.35]{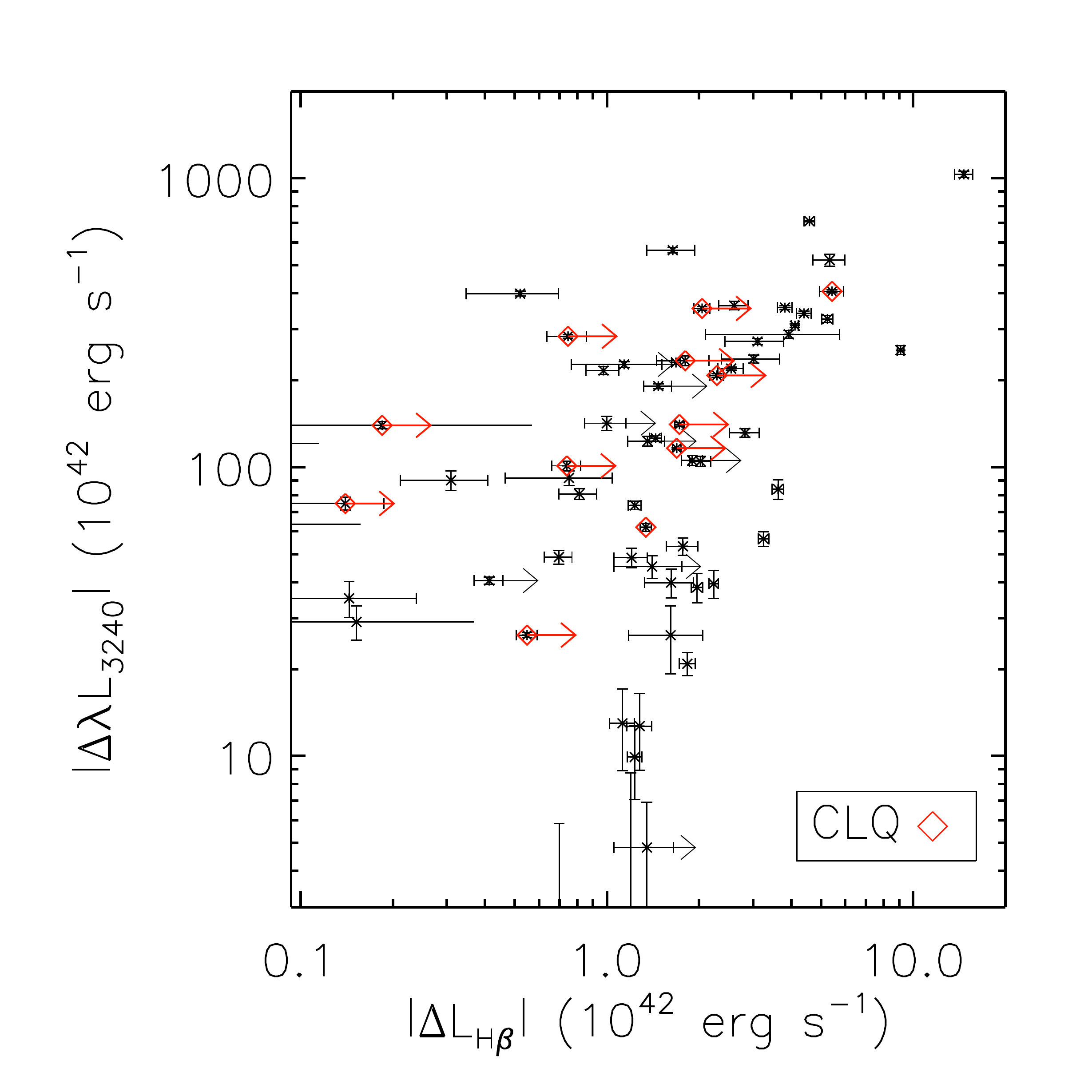}
\includegraphics[scale=.35]{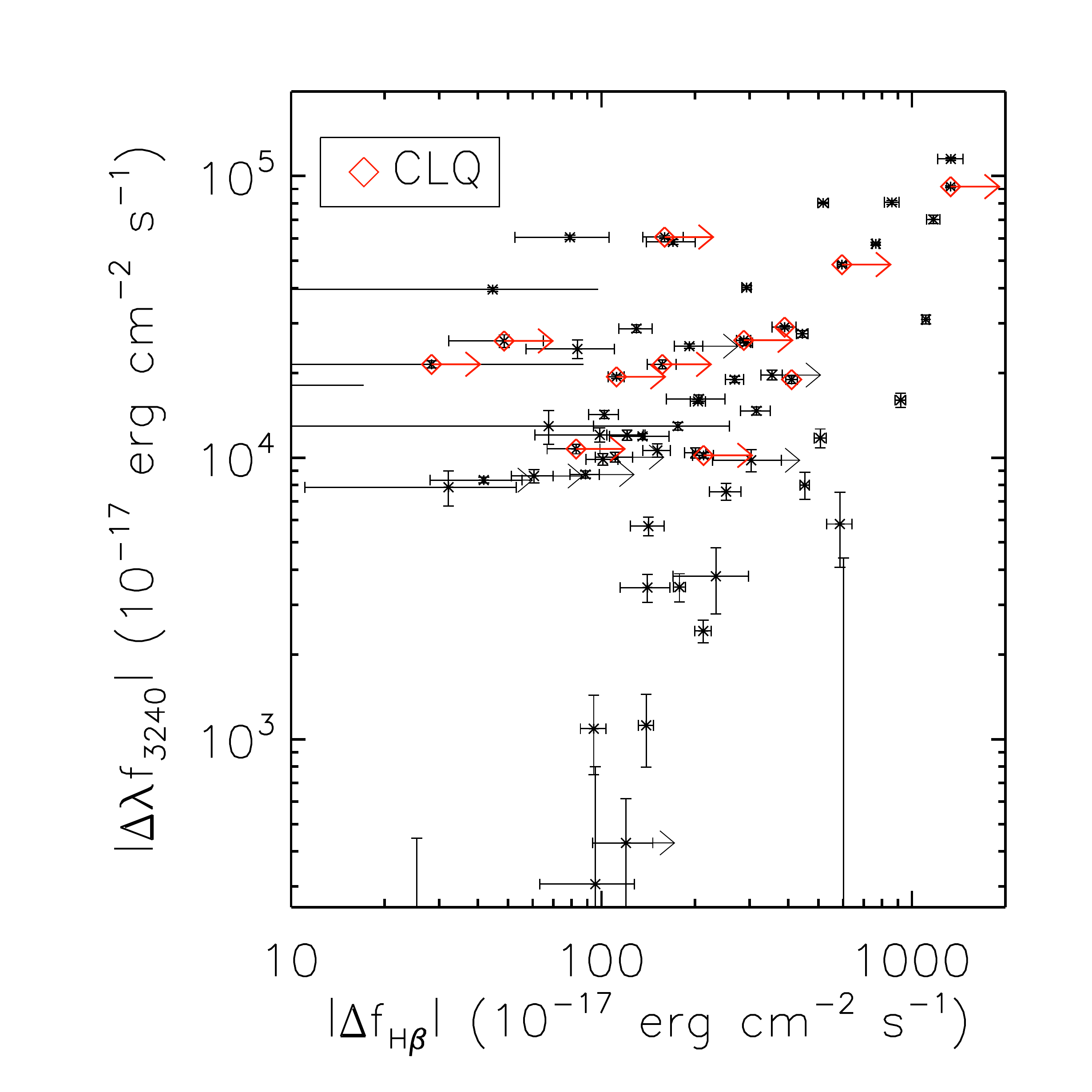}}
\caption{\footnotesize{
Absolute value of the 3240\AA\ continuum and H$\beta$
change in luminosity (\emph{left panel}) and flux (\emph{right panel})
for all analyzed CLQ 
candidates with a median dim-state S/N near H$\beta$ of $>5$. 
 The CLQs, defined as having absent broad H$\beta$ at one epoch
by at least $N_{\sigma}({\rm H}\beta)>3$,  are over-plotted as red
diamonds. Red horizontal arrows indicate lower limits on CLQs,
corresponding to upper limits on the amount of  H$\beta$ flux.  
}\label{fig:qsfit}}
\end{figure}

\begin{figure}[h!]
\centerline{
\includegraphics[scale=.4]{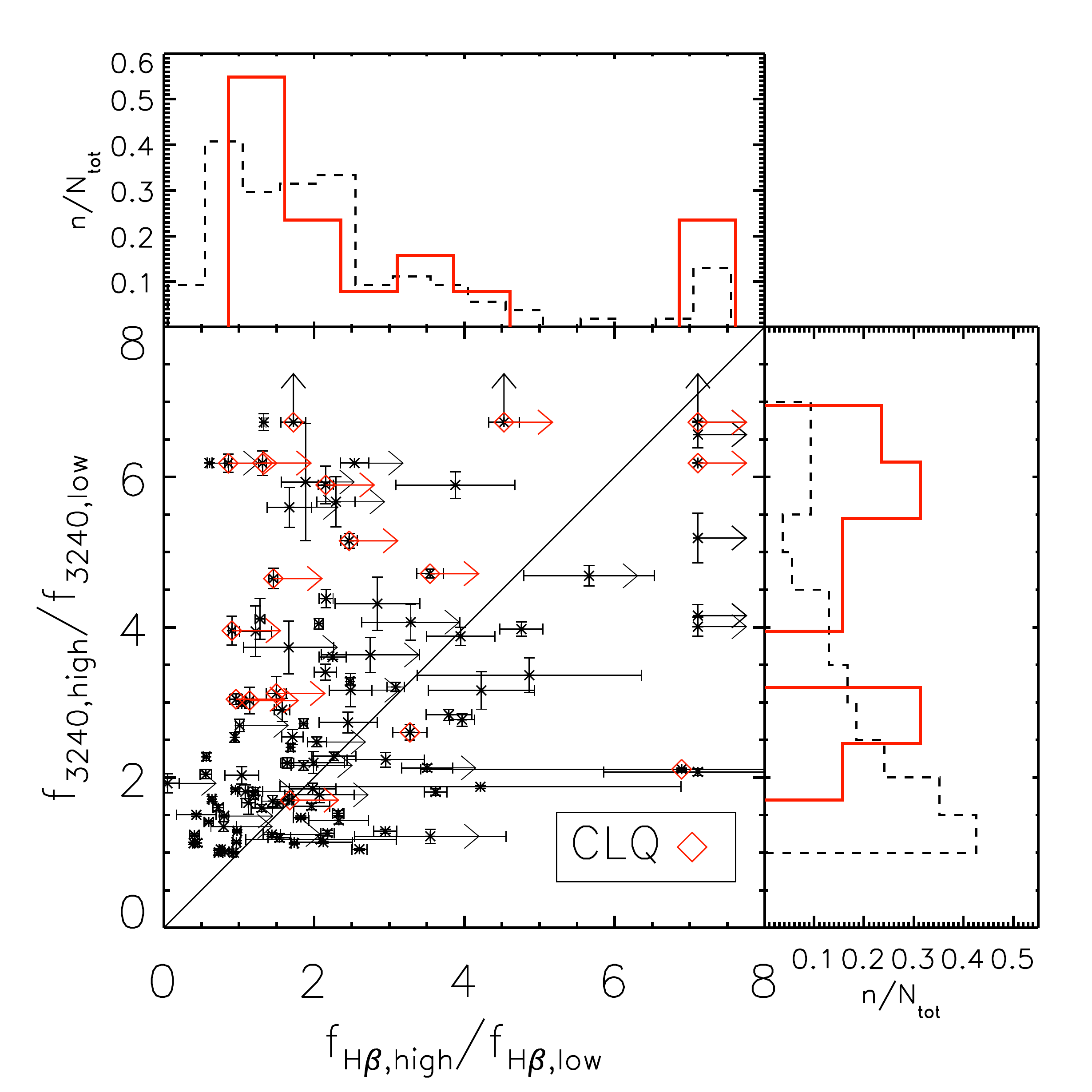}
\includegraphics[scale=.3]{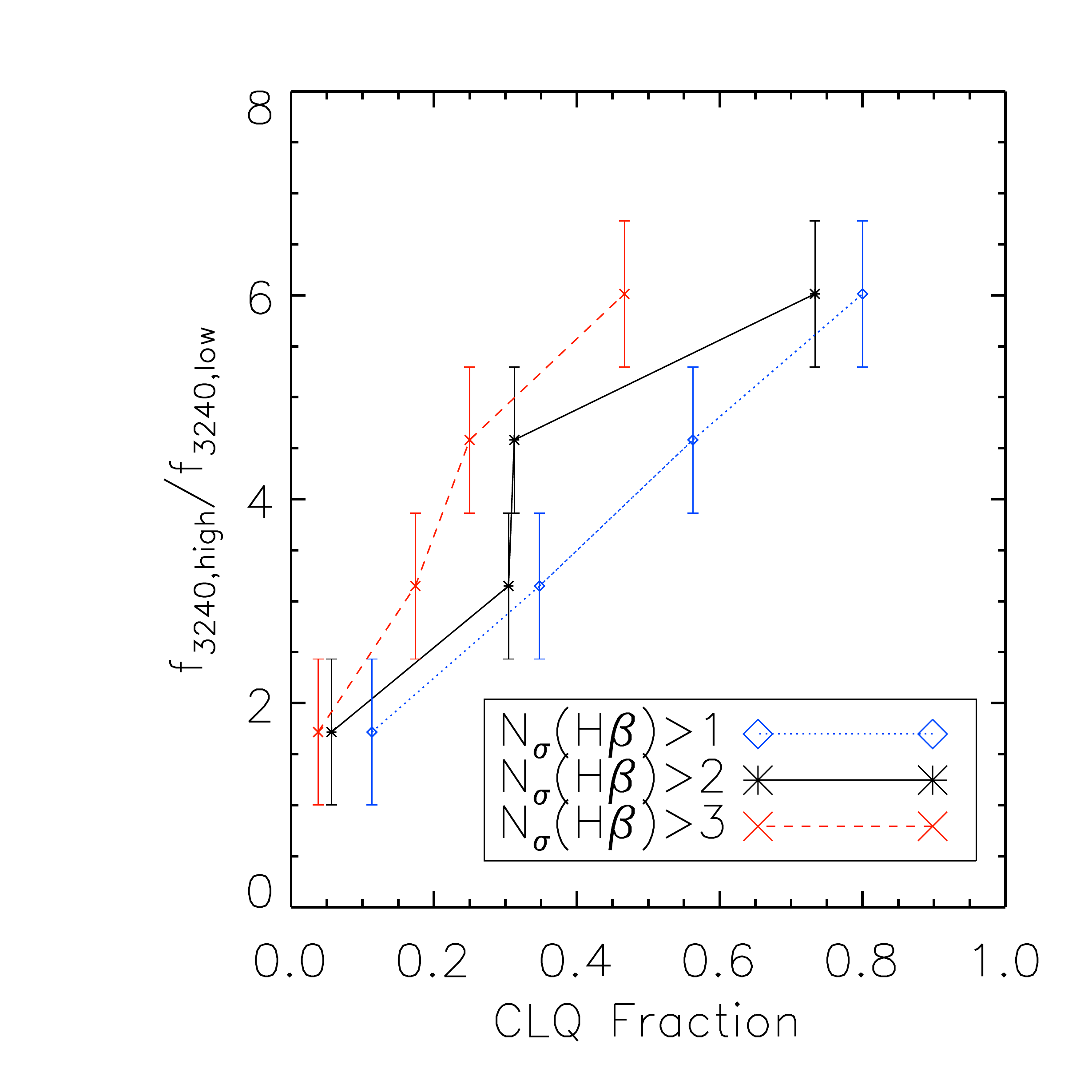}}
\caption{\footnotesize{
\emph{Left:}  Flux ratio at the 3240\AA\ continuum versus the
H$\beta$ flux ratio for CLQ candidates. The CLQs  are over-plotted as
red diamonds (and arrows) as in Figure~\ref{fig:qsfit}.  The ``high''
and ``low'' H$\beta$ 
fluxes are labeled according to the continuum level, so the ratios may
sometimes be $<1$  if the continua are similar, possibly due to
slightly inaccurate scaling of the follow-up spectrum. Black arrows
indicate locations of objects that fell outside the plotted range due
to small low-state flux levels. 
The line with unit slope is the expectation for a linear response of
BEL flux to the continuum variability.
The marginal distributions are shown on either axis, where CLQs are
shown in red and the overall sample is the black dashed histogram.  
\emph{Right:} CLQ fraction as a function of continuum flux ratio for
three different thresholds in $N_{\sigma}({\rm H}\beta)$. 
}\label{fig:qsfitfrac}}
\end{figure}

\begin{figure}[h!]
\centerline{
\includegraphics[scale=.4]{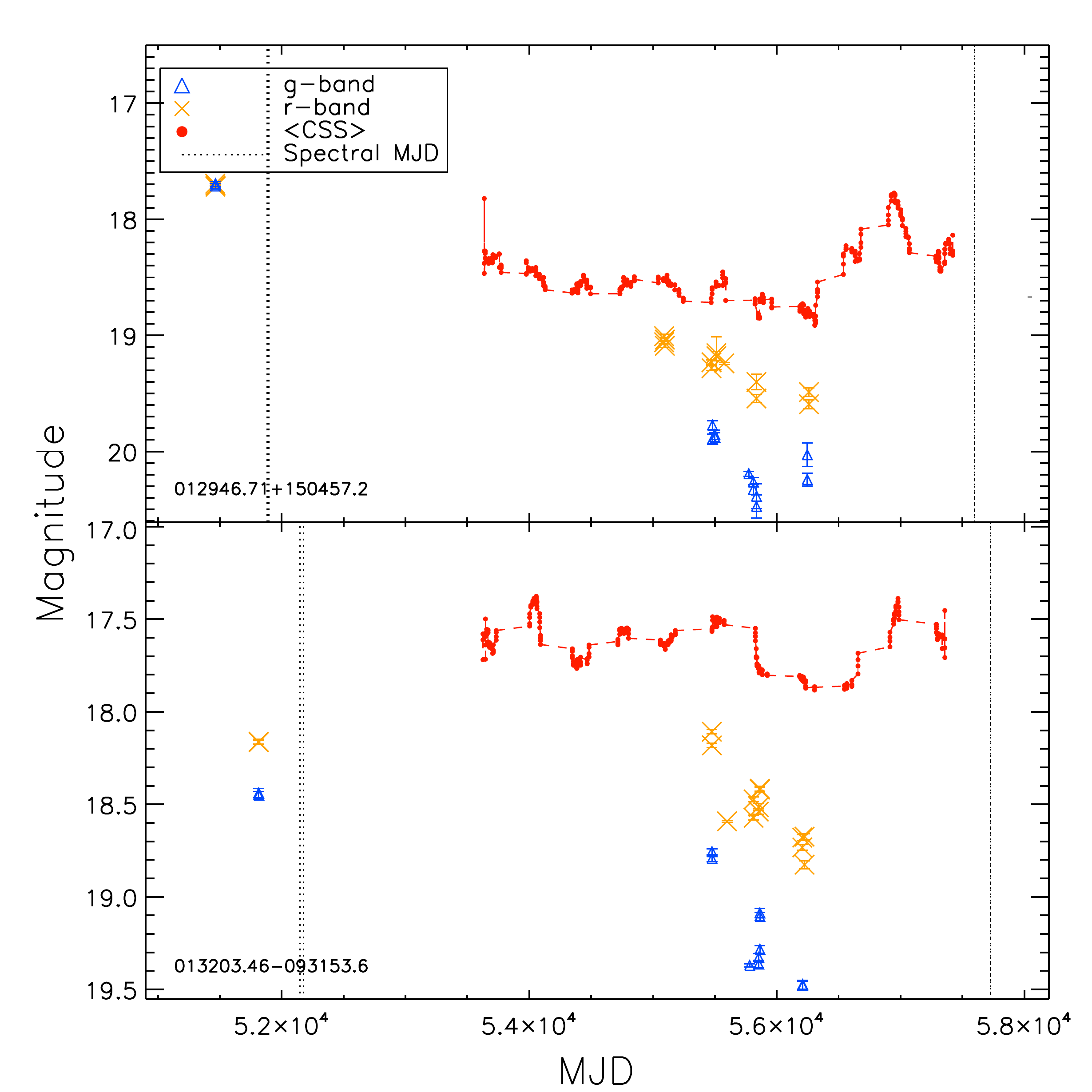}
\includegraphics[scale=.4]{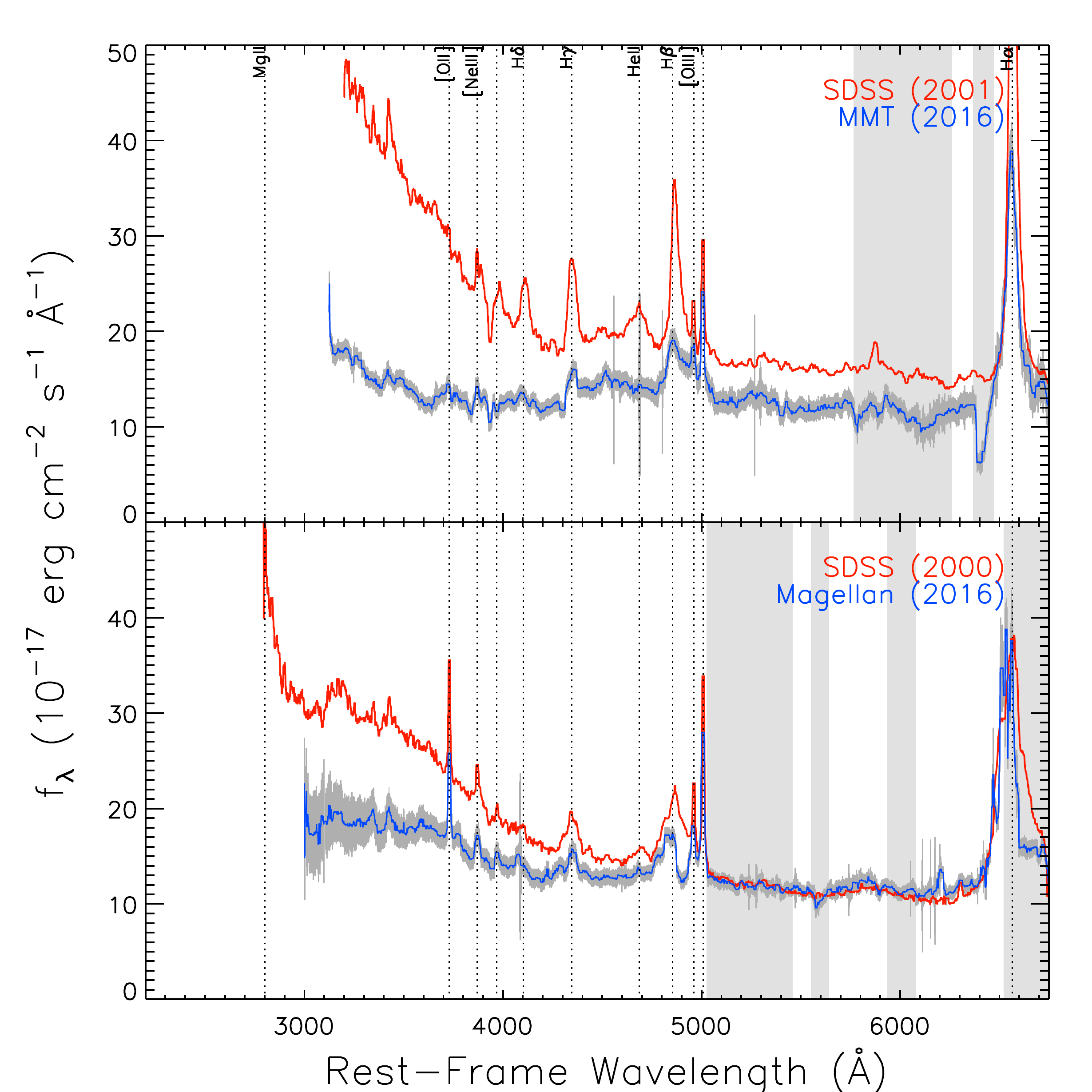}}
\caption{\footnotesize{
    {\it Left:} SDSS, PS1, and CRTS light curves for
    CLQ candidates J013203 (top) and J012946  (bottom), observed Dec.\
    4 2016 with MMT and 28 July 2016 with Magellan, respectively.  
    {\it Right:} Repeat spectra for each source in the adjacent left
    panel. J013203 shows a disappearance of \ion{He}{2} (see
    \S\,\ref{sec:He2}).  J012946 shows an remarkable
    disappearance of the red half of the Balmer BELs
     with $N_{\sigma}({\rm H}\beta)=6$  in
    response to a relatively large continuum change (\S\,\ref{sec:asym}). 
}\label{fig:obj_interest}}
\end{figure}

The following sources displayed asymmetric, boxy, or velocity-shifted
broad Balmer emission lines.  These are rare features seen usually in
low-luminosity quasars, and may be interesting since
they may indicate a rotating disk \citep[e.g.,][]{str03}, or a
supermassive black hole binary \citep{runnoe17}. In general, where
detectable, the H$\beta$ and H$\alpha$ profiles behaved similarly, but
the \ion{Mg}{2} profile
did not behave in the same way as the Balmer line profiles.  
\begin{itemize}
\item J141213.61+021202.1 (Figure~\ref{fig:clqA1}):
This object has lost the broad H$\beta$ component by a factor of 1.43
in flux (although at a
$<2\sigma$ level per spectral pixel), with a relatively small continuum
change of a factor 1.24.  In the earlier (bright) state, the broad
component was significantly \emph{blueshifted} relative to the narrow
component. 
\item J220537.71-071114.5 (Figure~\ref{fig:clq1}): 
This CLQ exhibits a \emph{redshifted} broad H$\beta$ component in the
bright state, before it disappears completely.  It also displays a
 down-turn of the UV part of the spectrum, possibly similar to the change
seen in J1100-0053 \citep{ross18}.  
\item J012946.71+150457.2 (Figure \ref{fig:obj_interest}):
This source maintained a blue contribution to the broad Balmer lines
among both (bright and dim) states, but loses the \emph{red half} of
the BEL in the later dim state. This behavior is very rare among AGN.
The strong change in BEL luminosity is
evident in Figure~\ref{fig:qsfit}, where the luminosity change is
nearly $|\Delta L_{H\beta}| = 9\times 10^{42}$ erg s$^{-1}$.
\item J143455.31+572345.0 (Figure \ref{fig:clq1}):
This object is listed as a disk-emitter (DE) quasar in \citet{str03}
and therefore is targeted as a DE QSO by the TDSS
few-epoch-spectroscopy program \citep[described in][]{mac18}. 
\item  There are additional examples of boxy or asymmetric Balmer
  profiles in our sample, e.g., in \S\ref{sec:flicker}, we described the CLQ J225240.37+010958.7
which has a noticeably extended red wing in the H$\beta$ profile.
The object J224829.47+144418.0 shown in Figure~\ref{fig:clqA1} shows a very
similar profile.  
A more detailed analysis is needed to
  classify all of them, which is outside the scope of this work.  
 In particular, we are unable to determine whether boxy profiles are more
or less frequent among the CLQ population because we lack a good
  benchmark sample to compare with, and the line profiles of CLQs may
  change shape as they dim which would not generally be 
  detectable given the dim state S/N typical here.
  A time-resolved transition between states would be
  especially useful in searching for emerging boxy profiles in
  dimming CLQs.  We defer further discussion of profile shapes to
   a future paper.

\end{itemize}

\floattable                                                                                                                            
\begin{deluxetable}{ccccccccccccccccc}                                                                                               
\rotate
\tablecaption{CLQ candidates\label{tab:clqcans}}                                                                                       
\tablecolumns{20}                                                                                                                      
\tablewidth{0pt}                                                                                                                       
\tablehead{                                                                                                                            
 \colhead{SDSSJID } & \colhead{ $z$} & \colhead{ morph. } & \colhead{ Phot.} &          
\colhead{ $g_1$} & \colhead{ $\sigma_1$ } & \colhead{Phot.} & \colhead{  $g_2$}                                                        
 & \colhead{ $\sigma_2$ } & \colhead{  Spec.} & \colhead{  MJD} &                                                                      
\colhead{  $g_{\rm PS1}$} & \colhead{  $\sigma_{\rm PS1}$ } & \colhead{ Spec. } & \colhead{Facility} & \colhead{CLQ} & \colhead{$N_{\sigma}$} \\
 \colhead{  } & \colhead{ } & \colhead{ flag } & \colhead{  MJD$_1$ } &                                   
\colhead{ (mag)} & \colhead{ (mag) } & \colhead{ MJD$_2$ } & \colhead{  (mag)}                                                         
 & \colhead{ (mag) } & \colhead{  MJD$_1$} & \colhead{ (PS1)} &                                                                        
\colhead{  (mag)} & \colhead{(mag) } & \colhead{MJD$_2$ } & \colhead{ } & \colhead{ by VI?} & $({\rm H}\beta)$\\                                       
}                                                                                                                                      
\startdata                                                                                                                             
000116.00$+$141123.0  &0.404  &0  &52170  &18.696  &0.021  &56588  &21.009  &0.071  &52235  &56588  & 21.009  &  0.071  &57989  &MMT       &0  &  2.6\\
000904.54$-$103428.7  &0.241  &0  &51814  &17.795  &0.028  &55860  &19.123  &0.022  &52141  &55860  & 19.123  &  0.022  &58367  &MMT       &1  &  8.2\\
001113.46$-$110023.5  &0.495  &1  &51865  &19.885  &0.031  &55829  &21.308  &0.112  &52141  &55834  & 21.233  &  0.113  &57989  &MMT       &1  &  2.3\\
001206.25$-$094536.3  &0.566  &0  &51865  &19.061  &0.049  &56218  &21.638  &0.148  &52141  &56218  & 21.638  &  0.148  &57597  &Magellan  &1  &  2.5\\
001502.38$-$094439.1  &0.336  &1  &51865  &18.877  &0.022  &56214  &20.349  &0.073  &52138  &56218  & 20.283  &  0.055  &57595  &Magellan  &0  &  2.0\\
002311.06$+$003517.5  &0.422  &0  &51081  &19.584  &0.024  &55449  &18.079  &0.032  &55480  &56206  & 18.728  &  0.016  &58037  &MMT       &1  &  2.4\\
002450.50$+$003447.7  &0.524  &1  &52522  &18.766  &0.030  &55449  &19.921  &0.133  &52203  &56206  & 19.814  &  0.029  &57989  &MMT       &0  &  0.4\\
002627.89$-$101020.5  &0.718  &0  &51814  &19.534  &0.026  &55829  &20.611  &0.066  &52145  &55829  & 20.583  &  0.063  &57596  &Magellan  &0  &  0.7\\
002714.21$+$001203.7  &0.454  &0  &51081  &18.261  &0.007  &56206  &19.466  &0.024  &51782  &56206  & 19.466  &  0.024  &57726  &MMT       &0  &  2.0\\
004339.32$+$134436.5  &0.527  &0  &51464  &19.349  &0.022  &55499  &21.997  &0.146  &51879  &55829  & 21.248  &  0.104  &57597  &Magellan  &1  &  3.1\\
005244.14$+$142807.1  &0.652  &0  &51464  &19.254  &0.021  &55745  &20.478  &0.034  &51871  &55745  & 20.478  &  0.034  &57597  &Magellan  &0  &  0.7\\
012821.43$+$151956.4  &0.548  &0  &51464  &18.669  &0.039  &55810  &20.190  &0.032  &51893  &55810  & 20.190  &  0.032  &57597  &Magellan  &0  &  2.2\\
012946.71$+$150457.2  &0.365  &0  &51465  &17.700  &0.023  &55835  &20.474  &0.098  &51898  &56246  & 20.029  &  0.101  &57597  &Magellan  &0  &  5.9\\
013203.46$-$093153.6  &0.190  &1  &51814  &18.439  &0.025  &56209  &19.479  &0.024  &52178  &56209  & 19.479  &  0.024  &57726  &MMT       &0  &  5.7\\
013458.36$-$091435.4  &0.443  &0  &51814  &18.605  &0.022  &56209  &19.985  &0.036  &52178  &56209  & 19.970  &  0.035  &57989  &MMT       &1  &  5.5\\
015957.64$+$003310.4  &0.312  &1  &51819  &19.122  &0.026  &54424  &20.284  &0.049  &51871  &55863  & 19.942  &  0.044  &55201  &BOSS  &1  & -\\
022556.07$+$003026.7  &0.504  &1  &52522  &19.974  &0.029  &55508  &21.783  &0.134  &52944  &56214  & 20.710  &  0.178  &55445  &BOSS  &1  & -\\
022652.24$-$003916.5  &0.625  &0  &52288  &20.252  &0.023  &55932  &22.002  &0.089  &52641  &56214  & 21.676  &  0.186  &56577  &BOSS  &1  & -\\
025428.83$-$004834.9  &0.810  &0  &52585  &19.759  &0.027  &56214  &20.831  &0.087  &52614  &56214  & 20.831  &  0.087  &  N/A  &N/A  &-  & -\\
\enddata                                                                                
\tablecomments{The full list of 262 candidates are available in the
  electronic version. Fields are marked with '-' if the repeat spectra
  were not analyzed (due to either data source or S/N) or not
  available. The third column lists the morphology flag (0 for point
  source, 1 if extended). }
\end{deluxetable}                                                                       

\floattable                                                                                          
\begin{deluxetable}{crrrrrrrr}                                                                       
\rotate
\tablecaption{Spectral Decomposition\label{tab:qsfit}}                                               
\tablecolumns{9}                                                                                     
\tablewidth{0pt}                                                                                     
\tablehead{                                                                                          
\colhead{SDSSJID } & \colhead{ $L_{\rm 3240,SDSS} $}  &  \colhead{Error in $L_{\rm 3240,SDSS} $} &   
\colhead{ $L_{\rm H\beta,SDSS} $ } & \colhead{Error in $L_{\rm H\beta,SDSS} $ } &                    
\colhead{ $L_{\rm 3240,2} $} & \colhead{Error in  $L_{\rm 3240,2} $} &                               
\colhead{ $L_{\rm H\beta,2} $ } & \colhead{Error in $L_{\rm H\beta,2} $} \\                          
\colhead{ } & \colhead{ $ (10^{42} {\rm ~erg~s}^{-1})$}  &  \colhead{$(10^{42} {\rm ~erg~s}^{-1})$} &
\colhead{ $(10^{42} {\rm ~erg~s}^{-1})$ } & \colhead{ $(10^{42} {\rm ~erg~s}^{-1})$} &               
\colhead{ $(10^{42} {\rm ~erg~s}^{-1})$}  & \colhead{ $(10^{42} {\rm ~erg~s}^{-1})$} &               
\colhead{ $(10^{42} {\rm ~erg~s}^{-1})$ } & \colhead{ $(10^{42} {\rm ~erg~s}^{-1})$} \\              
}                                                                                                    
\startdata                                                                                           
J000116.00$+$141123.0  &  509.9  &  7.2  &$ 7.02  $&  0.22  &  184.1  &  5.1  &$ 1.77$&  0.05  \\
J000904.54$-$103428.7  &  341.5  &  2.0  &$ 4.27  $&  0.05  &  201.0  &  0.4  &$< 2.55$&  0.03  \\
J001113.46$-$110023.5  &  244.5  &  7.3  &$ 2.19  $&  0.21  &   41.5  &  0.1  &$ 0.56$&  0.10  \\
J001206.25$-$094536.3  &  644.5  &  5.4  &$ 4.08  $&  0.28  &  158.4  &  9.2  &$< 1.24$&  0.23  \\
J001502.38$-$094439.1  &  133.2  &  3.0  &$ 1.95  $&  0.08  &   52.4  &  1.7  &$ 1.14$&  0.08  \\
J002311.06$+$003517.5  &  936.3  &  3.6  &$10.02  $&  0.19  &  378.5  &  9.8  &$< 4.92$&  0.29  \\
J002450.50$+$003447.7  &  691.5  &  5.0  &$ 4.60  $&  0.28  &  464.8  &  1.7  &$< 5.74$&  0.25  \\
J002627.89$-$101020.5  & 1040.9  & 20.8  &$ 9.96  $&  0.91  &  884.6  & 37.7  &$ 4.76$&  2.24  \\
J002714.21$+$001203.7  &  579.9  &  7.6  &$ 7.47  $&  0.54  &  816.8  &  2.0  &$ 4.45$&  0.35  \\
J004339.32$+$134436.5  &  215.6  &  5.6  &$ 6.60  $&  0.30  &    8.3  &  0.0  &$< 0.27$&  0.00  \\
J005244.14$+$142807.1  &  590.4  & 17.8  &$ 6.95  $&  0.47  &  288.8  &  0.3  &$12.45$&  1.50  \\
J012821.43$+$151956.4  &  666.3  &  8.8  &$ 6.66  $&  0.25  &  303.8  &  8.6  &$ 4.06$&  0.14  \\
J012946.71$+$150457.2  &  568.4  &  6.8  &$12.57  $&  0.19  &  314.4  &  5.7  &$ 3.47$&  0.14  \\
J013203.46$-$093153.6  &  139.5  &  0.5  &$ 1.47  $&  0.03  &   76.0  &  0.9  &$ 1.54$&  0.08  \\
J013458.36$-$091435.4  &  771.6  &  4.7  &$ 6.35  $&  0.13  &  365.6  &  2.0  &$ 0.92$&  0.46  \\
J025505.68$+$002522.9  &  744.5  &  1.3  &$ 7.52  $&  0.09  &  181.1  & 12.0  &$ 5.88$&  0.28  \\
J025606.03$+$001634.8  &  316.3  &  8.7  &$ 3.47  $&  0.34  &  503.8  &  6.4  &$ 4.52$&  0.21  \\
J025619.01$-$004501.3  & 1594.1  & 23.2  &$19.87  $&  0.93  &  562.3  & 12.1  &$ 5.24$&  0.34  \\
J033702.66$+$010627.5  &  108.6  &  4.6  &$ 1.79  $&  0.14  &  343.1  & 22.9  &$ 7.56$&  1.11  \\
\enddata                                                                                
\tablecomments{The subscript ``SDSS'' denotes the earlier epoch, SDSS spectrum, and the subscript ``2'' denotes the follow-up spectrum. The full list of 109 fits are available in the electronic version.}
\end{deluxetable}

\subsection{Distribution of Eddington Ratios}
\begin{figure}[h!]
\centering
\includegraphics[scale=.7]{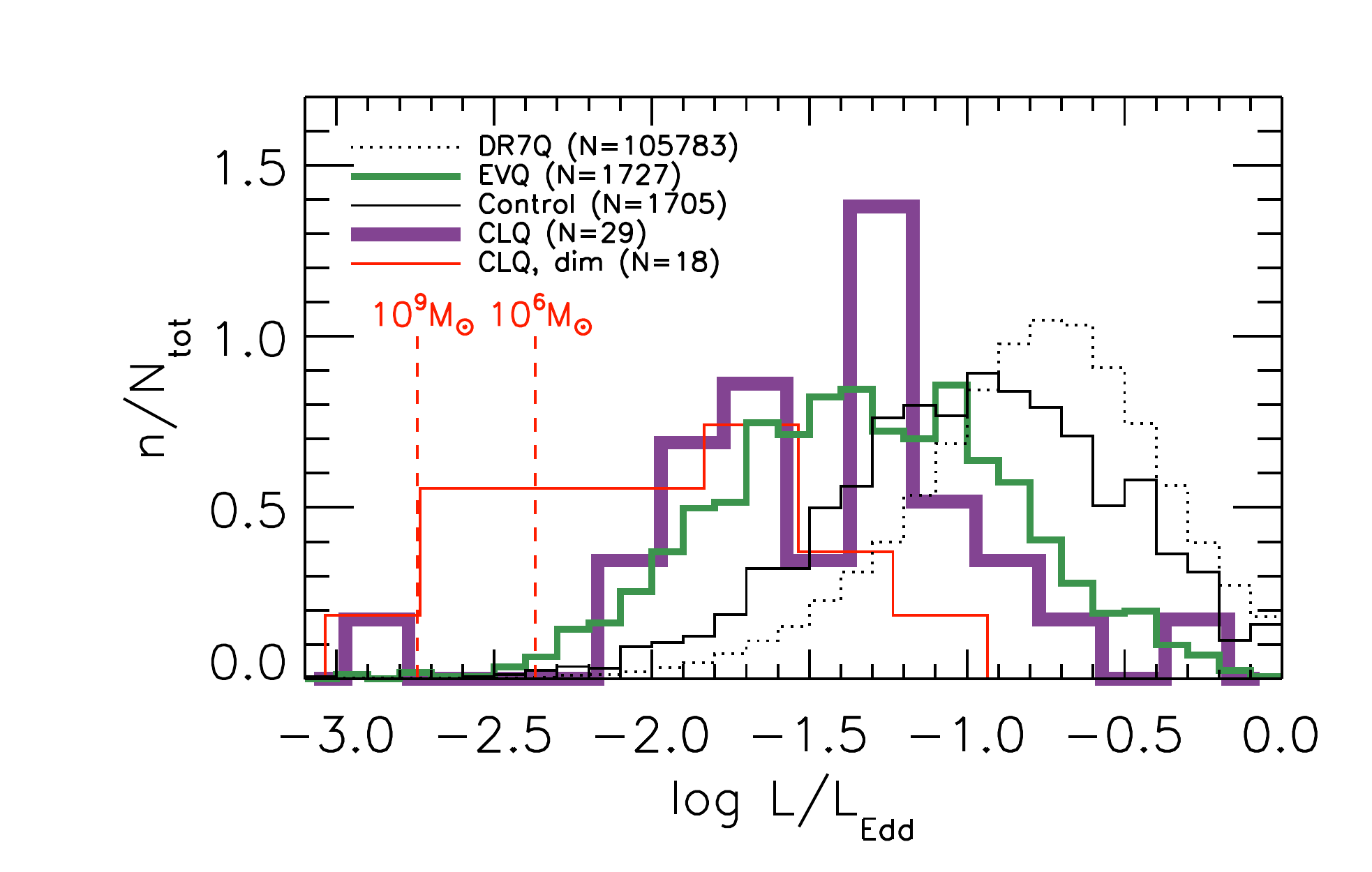}
\includegraphics[scale=.7]{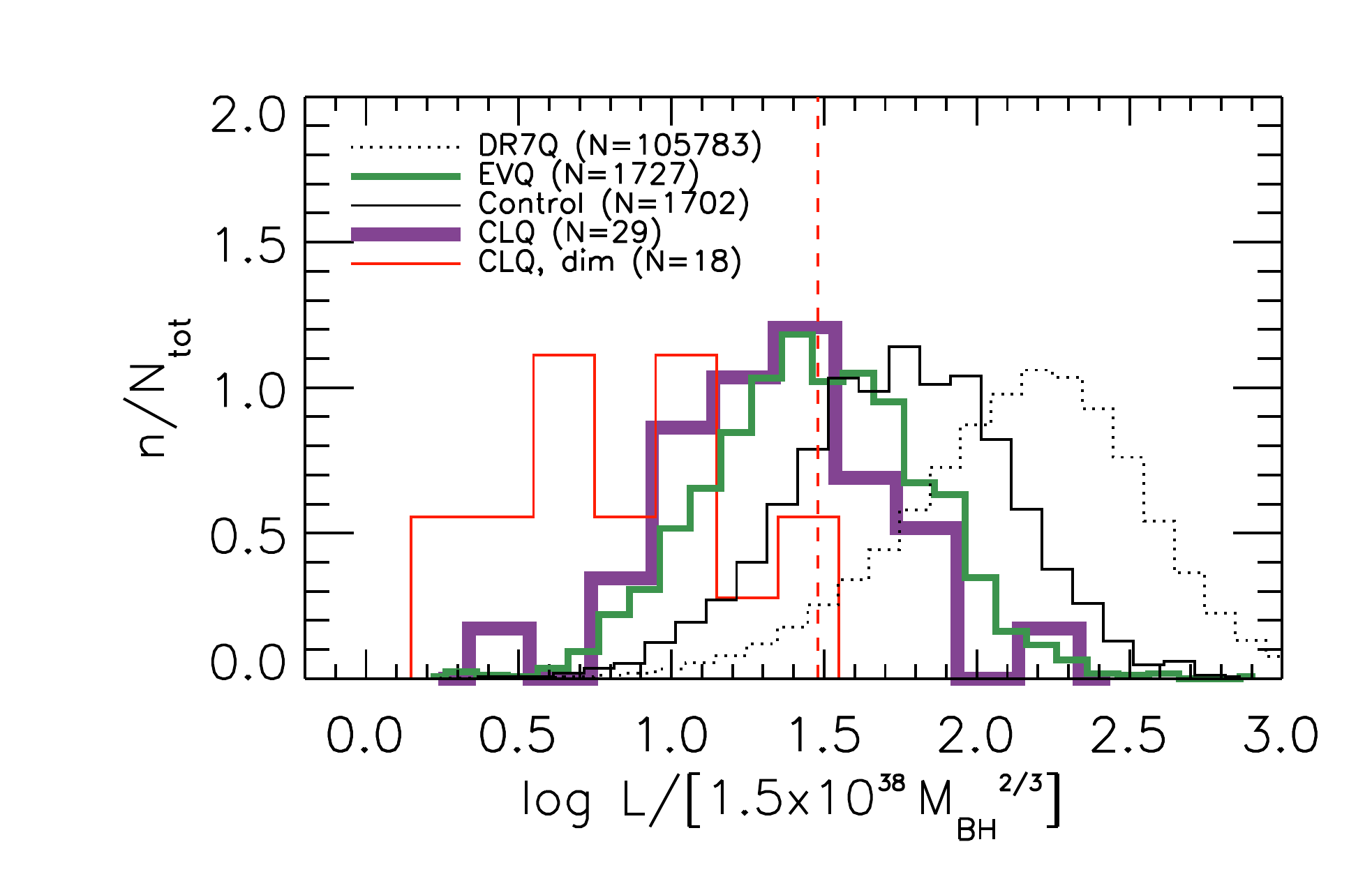}
\caption{    \emph{Top:}  The distribution of Eddington ratio for $z<0.83$ extremely variable
  quasars in SDSS / PS1 (EVQs, our parent sample; in green),  a
  less variable control sample matched in redshift and luminosity
  (in solid black), the full DR7 sample (dotted), and DR7 quasars with confirmed BEL
  changes (CLQs; in purple). 
  The thin red histogram shows the distribution
of the estimated dim state values for CLQs that were analyzed in \S\,\ref{sec:qsfit}.
Each histogram has been normalized to have unit area. Red vertical dashed lines indicate the predicted
critical values from \citet{nic00} above which BELs should be observable, for two different SMBH masses.
\emph{Bottom:} As in top panel but for the quantity  $\log (L_{bol} / [1.5\times
10^{38} M_{BH}^{2/3}])$, similar to the Eddington ratio,  where the BH
mass is in units of solar masses (this particular way of expressing
this quantity was chosen for simplicity). 
The quantity $L_{bol} / M_{BH}^{2/3}$ is the critical parameter in
the disk-wind model of \cite{eli09} and \cite{eli14} that determines whether or not a
BLR can form.  
The red vertical dashed line indicates the predicted
critical value above which BELs should be observable in this model. 
\label{fig:eddhist}} 
 \end{figure}

The Eddington ratio is an important physical quantity that has been
found to drive many features of quasar spectra and time variability. 
To test for any trend with Eddington ratio, we follow the procedure in
\citet{rum17} of constructing a less variable (non-EVQ) control sample 
and comparing the Eddington ratio distributions.  In particular,
  for each  quasar in the highly variable ($\Delta g > 1$~mag, $\Delta r >
  0.5$~mag) sample, we select a quasar at similar redshift and
  luminosity that does not meet the variability threshold, and form a
  control sample.  Our results
indicate that CLQs are indeed at lower Eddington ratios than the 
control sample 
(Figure\ \ref{fig:eddhist}).  Values of $L/L_{\rm Edd}$ from \citet{she11} are
adopted for each source, so the distribution shown is mostly for the
bright state spectra, although a handful of values will be for the dim state
\citep[e.g., J233317 from][which has $\log{(L/L_{\rm Edd})} =
-2.97$]{mac15}. The 3240\AA\ continuum luminosity typically 
changes by a factor of four, so the bolometric luminosity (and
therefore Eddington ratio) will be significantly smaller for the dim
state spectra.  We estimate the Eddington ratios in the dim state by simply
dividing the DR7 values by the amount that the 3240\AA\ continuum luminosity
has dimmed, based on our spectral decomposition described in
\S\,\ref{sec:qsfit}.  The thin red histogram shows the distribution
of the estimated dim state values for CLQs that were analyzed in \S\,\ref{sec:qsfit}. 
Note that the bolometric luminosity may have dimmed by a larger
amount than the 3240\AA\ continuum, so the red histogram is only a rough
estimate of the true distribution of Eddington ratios in the dim
state.

In \citet{eli09} and \citet{eli14}, the quantity  $L_{bol} / M_{BH}^{2/3}$ was
found to be the critical parameter in
a disk-wind model that determines whether or not a
BLR can form.  The bottom panel of  Figure\ \ref{fig:eddhist} shows
the distribution of this quantity for the same subsamples as in the
top panel. The vertical red dashed line shows the critical value
that divides Type 1 and true Type 2 AGN.  

We discuss the implications of this analysis in \S\,\ref{sec:physics}.

\section{Discussion}
\label{sec:disc}

The two fundamental questions we can  address are:  a) what is the CLQ
rate among highly variable AGN, and b) is the loss or emergence of
BEL flux in CLQs due to a rare and different physical mechanism than
that causing variability in the overall population of quasars? 
We find a CLQ
confirmation rate among highly variable quasars of 20\%, dependent 
on our luck with the timing of follow-up observations for some objects.
 While the CLQ fraction among highly variable quasars
was estimated to be $>15$\% in \citet{mac15}, they took the ratio of CLQs to the
\emph{total} number of highly variable quasars (on timescales of $\sim$8 years
rest-frame), regardless of the timing of the SDSS and BOSS spectral
epochs.  For our follow-up, we generally only target those quasars
that are currently at least 1 mag brighter or dimmer compared to the
$g$-band magnitude near any previous spectral epochs.

We also 
find that strongly varying quasars often show a large contrast in
Balmer BELs between dim and bright states, so when defining CLQs simply based on
(dis)appearing broad H$\beta$, the resulting CLQ fraction is highly
dependent on the significance of the change, $N_{\sigma}({\rm
  H}\beta)$. In terms of the flux ratio, CLQs are among those
variables with the largest changes, with the CLQ fraction
increasing from 10\% to roughly half as the continuum flux ratio
between states at 3420\AA\ increases from 1.5 to 6.

Based on this analysis, CLQs are consistent with being simply the tail
of a continuous distribution of quasar variability episodes. However,
the sample would benefit from a much larger size, especially at the
highly variable tail, to address this question. We discuss the
possibilities in \S\ref{sec:future}. Our results indicate that CLQs in
the dim state have lost a substantial fraction of their continuum luminosity
compared to other quasars. It is clear that this aspect causes the BLR to react
significantly, possibly structurally, to the continuum change.

The ``flickering'' CLQs J225240.37+010958.7 and J002311.06+003517.5 (\S\,\ref{sec:flicker})
show that objects with substantial broad BEL variability appear to
repeat this behavior whenever the continuum varies significantly.
Similar persistent variability has also been seen in SDSS~J022556 \citep{mac15}
and Mkn 1018 \citep{hus16}. 
Furthermore, \citet{rum17} find longer characteristic timescales and larger
variability amplitudes, \emph{as measured from SDSS S82}, for the
sample of extremely variable quasars (EVQs) found in DES/SDSS data. This result
makes sense in view of the red noise variability power spectrum of
quasars, where the power index of $-2$ toward higher frequencies leads
to larger variability amplitudes on longer timescales. These findings
suggest that previously identified strong variability predicts the same
behavior in the future. It is possible  that a small fraction of the
CLQs presented here and elsewhere exhibit physically distinct
variability from the rest of quasar population, but that most of the
CLQs are actually just the extreme tail of regular quasar variability.
Here, we discuss the physical properties of CLQs on average, and leave
a detailed analysis for particular sources for future study.

\subsection{Physical Origin of CLQ Variability}
\label{sec:physics}

Studies have investigated whether disappearing BELs in the optical are
likely to originate from extrinsic or transient events  
\citep[e.g.][]{goo89,goo90,lam15,mer15,rua15,run15,gez17}. In
\citet{mac15}, we found that variable extinction by dust cannot 
account for BEL flux changes relative to the continuum in CLQs, nor the observed
timescales. In general the flares are longer-lived than the typical
TDE \citep{gui14}, and pre-existing narrow
emission lines in CLQ spectra are not expected for a single TDE outburst. 
The CLQs studied here are all $z<1$, and lack foreground spectral
features so their variability is unlikely to be the result 
of lensing by a foreground galaxy \citep[see, e.g.,][]{qui14}.
Furthermore, erratic variability and multiple flaring episodes are
evident in the light curves   
of CLQs, as opposed to single Paczy\'nski curves one might expect for
high-amplitude microlensing by a single foreground star in a faint
foreground galaxy \citep{bru17}. Therefore, extreme variability is
likely intrinsic to the quasar and not due to lensing.  

We are now able to specifically confirm a trend with Eddington-ratio
from our follow-up spectroscopy, and it is consistent with the well
known anti-correlation between Eddington ratio and variability
amplitude \citep[e.g.,][]{wil08,mac10}. We find that CLQs have lower
Eddington-ratios than a control sample matched in redshift and
luminosity, as suggested by \citet{rum17}. Since CLQs seem to occupy
this region of physical parameter space, as Rumbaugh et al.\ point out, BEL
(dis)appearance in general is not likely to be due to variable
obscuration, tidal disruption events (TDEs), or microlensing by
foreground stars, unless these events are strongly preferred in
quasars with lower Eddington ratio.

Interestingly, the CLQs are found near the critical luminosity below
which the BLR disappears \citep[][vertical line in bottom panel of
Figure~\ref{fig:eddhist}]{eli09}. This supports a picture where the 
accretion rate in CLQs is barely enough to
support a broad line region, assuming a disk-wind model for the BLR.
\citet{nic00} also assume a disk-wind model and derive a critical
threshold in $L_{\rm bol}/L_{\rm Edd}$ for BLR disappearance as a
function of black hole mass (vertical lines in top panel of
Figure~\ref{fig:eddhist}). Here, we assume a
maximum accretion efficiency of $\eta=0.06$ and a viscosity coefficient of
$\alpha=0.1$ as in \citet{nic00}. The CLQs begin to reach this threshold if we
estimate $L_{\rm bol}/L_{\rm Edd}$ in the dim state (conservatively
assuming the same SED as in the bright state).  
Assuming the variability is indeed due to an accretion rate
change, and that the estimated dim state values for CLQs as well as
the adopted values for $\eta$ and $\alpha$ are not too inaccurate,
our results indicate that the \citet{eli09} model for the disk-wind
BLR may provide a good description of CLQs, since the CLQ distribution
would need to be a  factor of 10 lower in  $L_{\rm bol}/L_{\rm Edd}$
before reaching the \citet{nic00} critical threshold between quasars
that do and do not have a BLR.

Understanding the various physical processes that occur at low
accretion rates is therefore highly relevant to the study of
changing-look variability.  A key question is how to connect the
observed timescales for CLQ transitions with theoretical timescales
for accretion disks. The expected timescale for a major accretion rate
change is the viscous timescale  \citep{kro99}. However, the expected
viscous timescale in the optical-emitting region of the disk is
orders of magnitude longer than what is observed for CLQ transitions
\citep[for a summary of the timescale problem, see][]{law12}. Various
solutions to the problem of observed transition timescales have been
proposed, some involving thermal instabilities, as we discuss
below. The results from \citet{hus16} on Mkn 1018 indeed support a
disk temperature change associated with the CLQ event, based on the
measured temperature in both states and corresponding luminosity
dependence. Additionally, the timescale is consistent with a thermal
timescale for an accretion disk. Other theoretical timescales include
the dynamical and sound-crossing time; the former is usually
too short to explain the observations, whereas the radial
sound-crossing time is usually too long \citep[e.g., see table in][]{law16}.
Either the instability (or instabilities)  must propagate throughout
the disk far faster than the viscous timescale, such as over a thermal
timescale, or our calculation of a viscous timescale is off by orders
of magnitude. 

State of the art simulations of the thermal instability in the
radiation-dominated region of accretion disks by \citet{jia16}
indicate that disks with lower-metallicity gas are more prone to
thermal instability, as the iron opacity bump is less effective as a
stabilizing mechanism. Therefore, a reliable way to measure
the metallicity in the inner regions of quasars may be relevant to the
interpretation of CLQs. 

A recent idea  that would explain the fast CLQ variability invokes a
cooling front due to a sudden change in torque applied by the magnetic
field at the inner-most stable circular orbit
\citep{ross18,ste18}. Another idea involves an  advection-dominated
accretion flow \citep[ADAF;][]{nod18}. In the latter scenario, the BLR 
disappearance is triggered by  a sudden drop in accretion rate,
leading to an ADAF in the inner disk; the consequence is a harder
spectral energy distribution. \citet{dex18} suggested that 
all quasar accretion disks are actually  magnetically elevated, 
leading to larger scale heights and shorter variability
timescales than expected in standard thin disk theory \citep{SS73}. 
 In this case, CLQs may simply be the tail of a continuous
distribution of quasar variability.

Multi-wavelength observations for more CLQs are needed, as the SED can
suggest certain physical environments. For example,
X-ray repeat spectroscopy for a handful of sources rules out variable
absorption as the cause of variability  \citep[][and references
therein]{lam15}, as well as do mid-IR light curves
\citep[e.g.,][]{ross18,ste18}. One of the best
studied sources, NGC2617 \citep{sha14}, has simultaneous UV--IR
monitoring along with X-rays that show reprocessed emission from a
central X-ray source with no signs of dust obscuration along the line
of sight to the central engine.
Through SED modeling, \citet{kub18} determined that X-ray
reprocessing becomes increasingly important at lower
accretion rates, but it cannot account for all the
optical-UV variability in SDSS quasars.

\subsection{Future Work}
\label{sec:future}

Our sample is dominated by dimming events (or ``turn-off'' CLQs)
because of the sample construction: since our parent sample is the
DR7 quasar catalog,
most have broad BELs in the early state.  Only two of the CLQs
presented here are transitioning to a bright state
(J105553.51+563434.4 and J214852.50$-$000554.7; Figure~\ref{fig:clqA1}). 
The turn-off CLQ frequency is likely balanced by turn-on events,
unless we expect the population to be evolving. A major piece of
future work is therefore to establish a suitable parent sample and
method for searching for turn-on events. The challenge in starting
from a galaxy catalog is how to produce reliable photometry for just
the (mildly active) nuclei of extended galaxies, as well as how to
efficiently sift through a sample an order of magnitude larger than
the quasar sample.  For a recent effort, see \citet{dra19}.

Another outstanding question is why the \ion{Mg}{2} BEL, formed at a
similar ionization energy as the Balmer lines, is generally less responsive to
strong changes in the continuum \citep[see also][]{cac15}.  
Given that reverberation mapping studies in the past have
typically focused on lower-redshift quasars, where  \ion{Mg}{2} is outside the
optical wavelength range \citep[although see][]{she16}, 
characterizing the response of the \ion{Mg}{2} line is an important
goal. We leave the \ion{Mg}{2} BEL variability measured from these follow-up spectra
for a future publication (Homan et al.\
2019).  

Having well-sampled light curves for quasars is essential for
determining the timescales associated with the transitions and for
selecting targets for follow-up. Therefore, future multi-band surveys
in the time domain such as the Large Synoptic Survey Telescope
\citep{ive08,lsst}, PS2 \citep{cha14}, Zwicky Transient Facility
\citep[ZTF,][]{ztf}, as well as, for brighter AGN, the ongoing All-Sky
Automated Survey for SuperNovae \citep[ASASSN;][]{sha14}, should
provide many interesting targets for contemporaneous spectroscopic follow-up. 
For example, ZTF commenced in March 2018 and is monitoring 15,000
deg$^2$ of the Northern Sky in $g$ and $r$ bands with a cadence of 3
nights down to a magnitude of 20.5 mag over a baseline of 3 years.
While we can predict which variability episodes are uncommon with 
respect to the overall quasar population \citep[e.g.,][]{gra17}, 
one issue is how to efficiently flag CLQs in photometric monitoring data
\emph{before} they change, so that we can spectroscopically
monitor them in the optical \emph{during} the transition.
Predictive modeling, where the expected variability is predicted
either for a given night or over a particular timeframe, could prove
useful here (see Graham et al., 2019, PASP). 

It is unclear whether the same physical processes are occurring
in/around the accretion disks in extremely variable quasars.  But
there are some hints of outliers and interesting objects; this problem
will also be easier to solve with larger datasets of repeat quasar
spectroscopy.  Samples have recently grown substantially in the
SDSS-IV TDSS \citep{mor15,mac18}, forming a pilot to even larger
spectroscopic surveys upcoming in SDSS-V \citep{kol17}.
A sample of newly discovered CLQs from SDSS-IV will be presented in MacLeod
et al.\ (2019).  The ideal survey would provide dense cadence
repeat spectroscopy of a large sample of quasars, e.g., via time
domain objective prism survey, or a many-band space mission such as
SPHEREx \citep{dor18} or the  Time-domain Spectroscopic Observatory\footnote{\tt https://pcos.gsfc.nasa.gov/physpag/probe/TSO-Probe-WhitePaper-submit.pdf}.   
One potential result from a large survey of repeat quasar spectroscopy
is to expand the sample size  of CLQs with multiple spectra before,
during, and after the transition.  With a larger sample size, we can 
detect and characterize discontinuities in the overall variability
distributions associated with CLQ behavior.

\acknowledgments

This material is based upon work supported by the National Science
Foundation under Grants No. AST-1715763 and AST-1715121. 
The work of DS was carried out at the Jet Propulsion
Laboratory, California Institute of Technology, under a contract
with NASA. We thank Marco Lam and Nigel Hambly for assistance with the PS1 DVO
database hosted at Edinburgh.  
We thank Peter Blanchard for assistance in carrying out Blue Channel
observations. 
CLM would like to acknowledge Zeljko Ivezic, Martin Elvis and Fabrizio Nicastro for
useful discussions regarding the Eddington ratio and disk-wind model.  
We thank an anonymous referee for their very helpful and thorough comments.

Funding for the SDSS and SDSS-II has been provided by the Alfred P. Sloan Foundation, the Participating Institutions, the National Science Foundation, the U.S. Department of Energy, the National Aeronautics and Space Administration, the Japanese Monbukagakusho, the Max Planck Society, and the Higher Education Funding Council for England. The SDSS Web Site is http://www.sdss.org/.

The SDSS is managed by the Astrophysical Research Consortium for the Participating Institutions. The Participating Institutions are the American Museum of Natural History, Astrophysical Institute Potsdam, University of Basel, University of Cambridge, Case Western Reserve University, University of Chicago, Drexel University, Fermilab, the Institute for Advanced Study, the Japan Participation Group, Johns Hopkins University, the Joint Institute for Nuclear Astrophysics, the Kavli Institute for Particle Astrophysics and Cosmology, the Korean Scientist Group, the Chinese Academy of Sciences (LAMOST), Los Alamos National Laboratory, the Max-Planck-Institute for Astronomy (MPIA), the Max-Planck-Institute for Astrophysics (MPA), New Mexico State University, Ohio State University, University of Pittsburgh, University of Portsmouth, Princeton University, the United States Naval Observatory, and the University of Washington.

Funding for SDSS-III has been provided by the Alfred P. Sloan Foundation, the Participating Institutions, the National Science Foundation, and the U.S. Department of Energy Office of Science. The SDSS-III web site is http://www.sdss3.org/.

SDSS-III is managed by the Astrophysical Research Consortium for the Participating Institutions of the SDSS-III Collaboration including the University of Arizona, the Brazilian Participation Group, Brookhaven National Laboratory, Carnegie Mellon University, University of Florida, the French Participation Group, the German Participation Group, Harvard University, the Instituto de Astrofisica de Canarias, the Michigan State/Notre Dame/JINA Participation Group, Johns Hopkins University, Lawrence Berkeley National Laboratory, Max Planck Institute for Astrophysics, Max Planck Institute for Extraterrestrial Physics, New Mexico State University, New York University, Ohio State University, Pennsylvania State University, University of Portsmouth, Princeton University, the Spanish Participation Group, University of Tokyo, University of Utah, Vanderbilt University, University of Virginia, University of Washington, and Yale University.

The Pan-STARRS1 Surveys (PS1) have been made possible through contributions of the Institute for Astronomy, the University of Hawaii, the Pan-STARRS Project Office, the Max-Planck Society and its participating institutes, the Max Planck Institute for Astronomy, Heidelberg and the Max Planck Institute for Extraterrestrial Physics, Garching, The Johns Hopkins University, Durham University, the University of Edinburgh, Queen's University Belfast, the Harvard-Smithsonian Center for Astrophysics, the Las Cumbres Observatory Global Telescope Network Incorporated, the National Central University of Taiwan, the Space Telescope Science Institute, the National Aeronautics and Space Administration under Grant No. NNX08AR22G issued through the Planetary Science Division of the NASA Science Mission Directorate, the National Science Foundation under Grant No. AST-1238877, the University of Maryland, and Eotvos Lorand University (ELTE).

The CSS survey is funded by the National Aeronautics and Space
Administration under Grant No. NNG05GF22G issued through the Science
Mission Directorate Near-Earth Objects Observations Program.  The CRTS
survey is supported by the U.S.~National Science Foundation under
grants AST-0909182 and AST-1313422.

\vspace{5mm}
\facilities{SDSS, Pan-STARRS, WHT, MMT, Magellan, CRTS, Palomar}

\software{pyDIS, QSfit, pyRAF, MySQL/CASJobs}

\appendix
\section{Less-significant CLQs and near-CLQs}
\label{sec:nearclqs}

We show in Figure~\ref{fig:clqA1} the visually-identified CLQs with relatively low significance, in the range $1<N_{\sigma}({\rm H}\beta)<3$. In Figure~\ref{fig:nearclqs}, we present some examples where the H$\beta$ line lost much of its broad component along with some dimming of the continuum, but not enough to classify it as a CLQ. 

\begin{figure}[h!]
  \subfloat[][Fig. A1$a$]{\centerline{
\includegraphics[scale=.4]{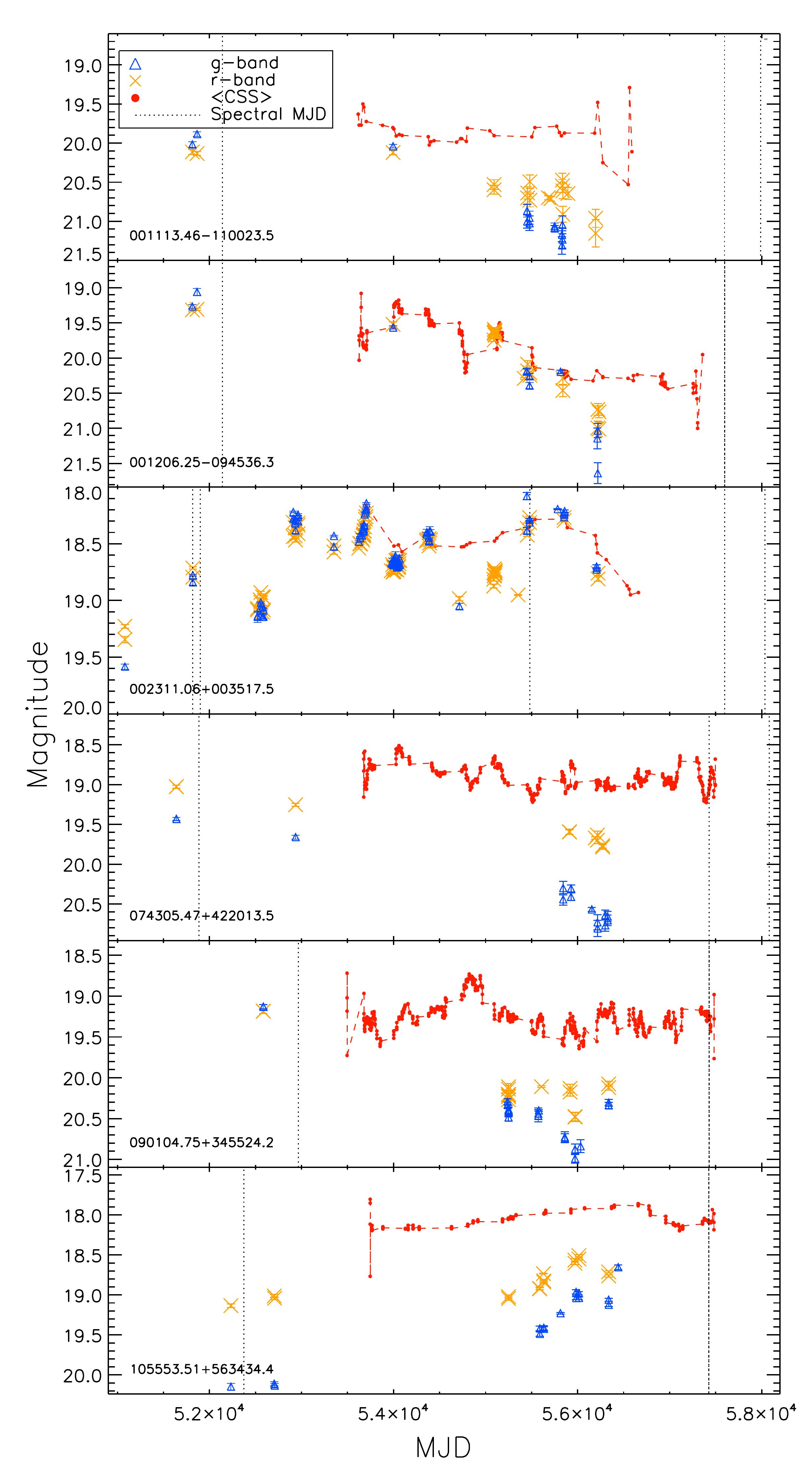}
\includegraphics[scale=.4]{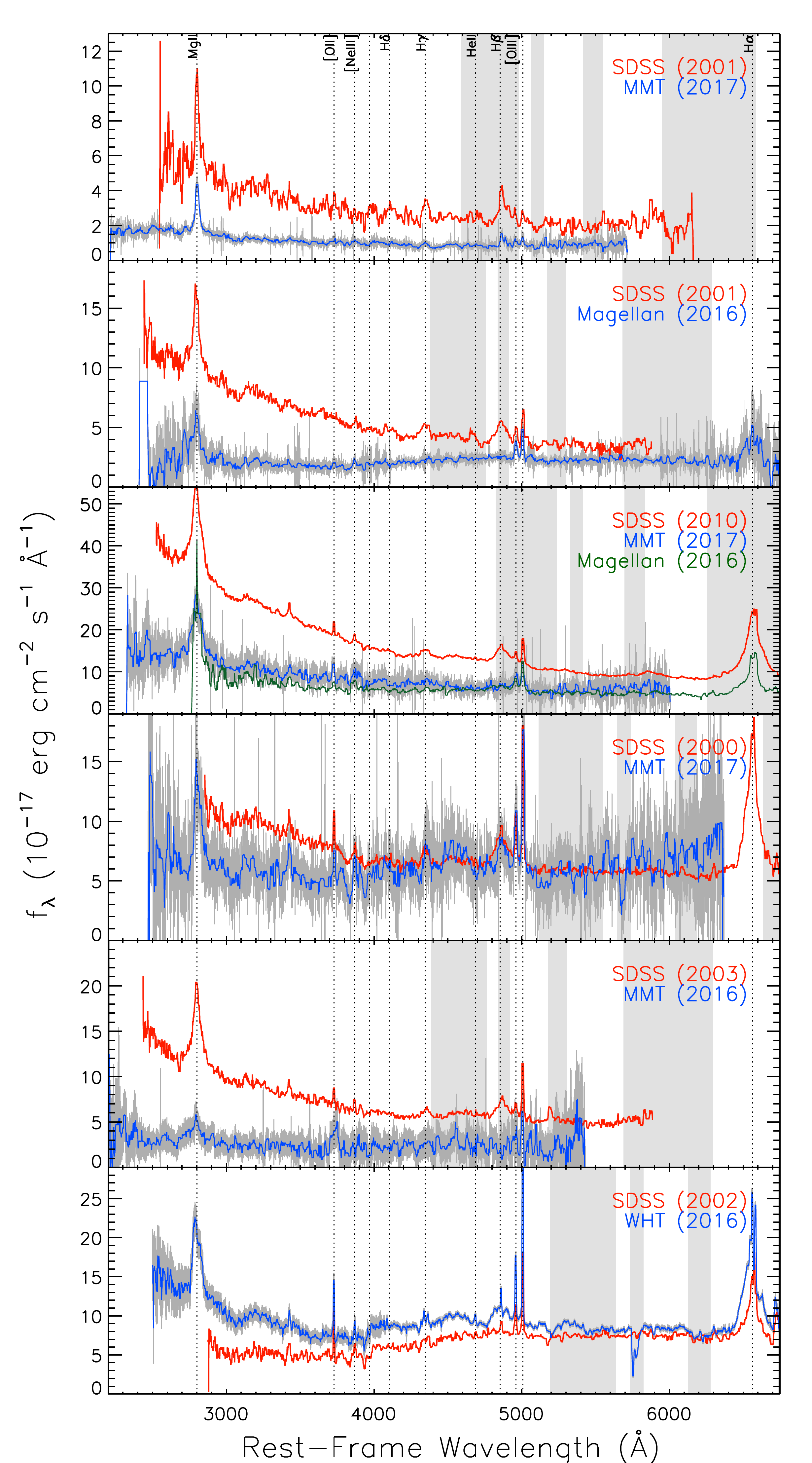}}}
\caption{\footnotesize{
As in Figure~\ref{fig:clq1}, but for lower-significance CLQs
 with $1<N_{\sigma}({\rm H}\beta)<3$.  
 Light curves and repeat spectra shown in R.A.\ order. 
 \emph{Left:}   SDSS and Pan-STARRS $g$-band photometry is shown
 as blue triangles. Red data points show archival photometry from CRTS.
 The existing spectroscopic epochs are indicated by the vertical lines. 
 \emph{Right:} Existing spectra for objects in the adjacent left panel (red is
 SDSS; blue or green is our follow-up), corrected for telluric
 absorption where needed. 
}\label{fig:clqA1}}
\end{figure}
\begin{figure}[h!]
  \ContinuedFloat
  \subfloat[][Fig. A1$b$]{\centerline{
\includegraphics[scale=.4]{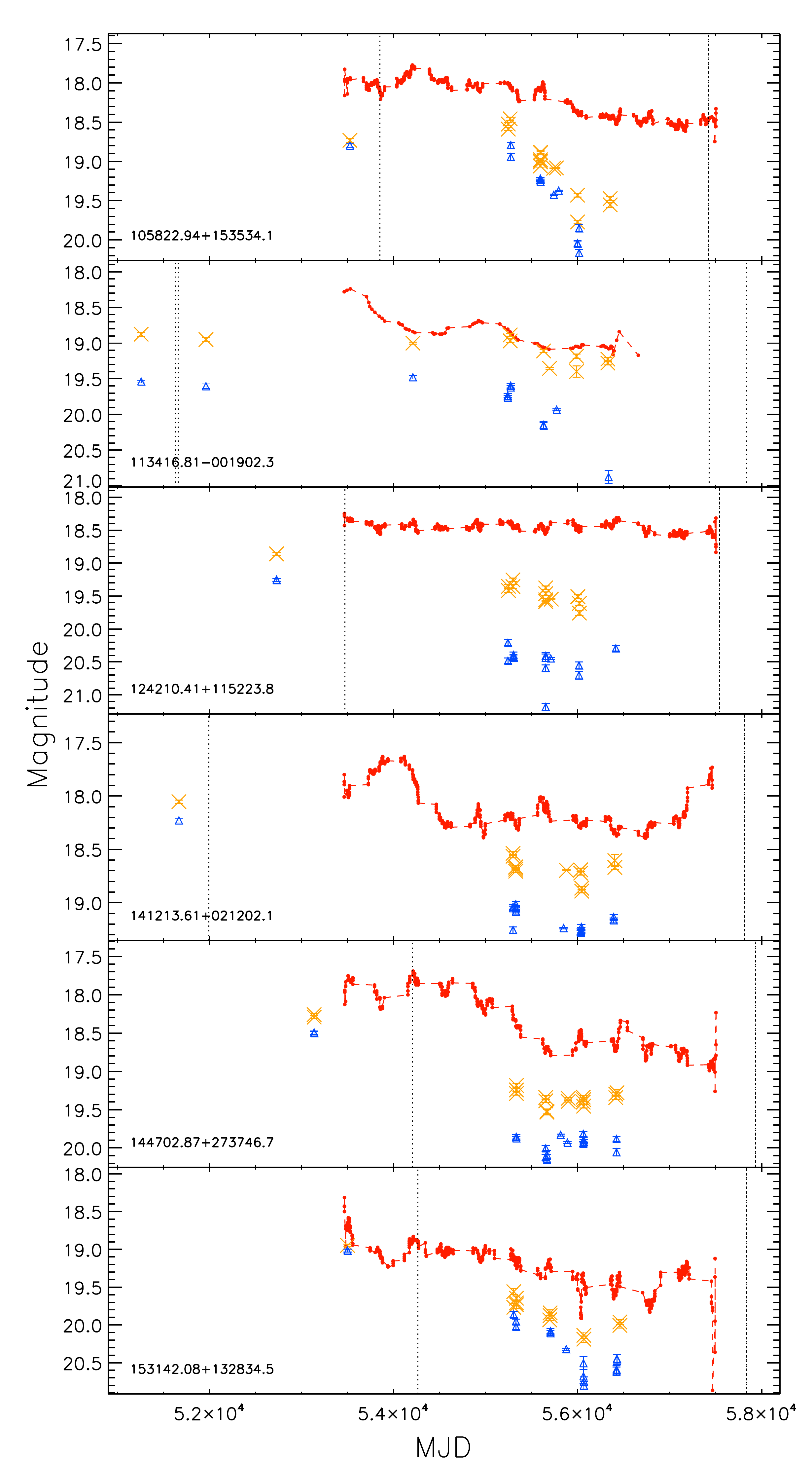}
\includegraphics[scale=.4]{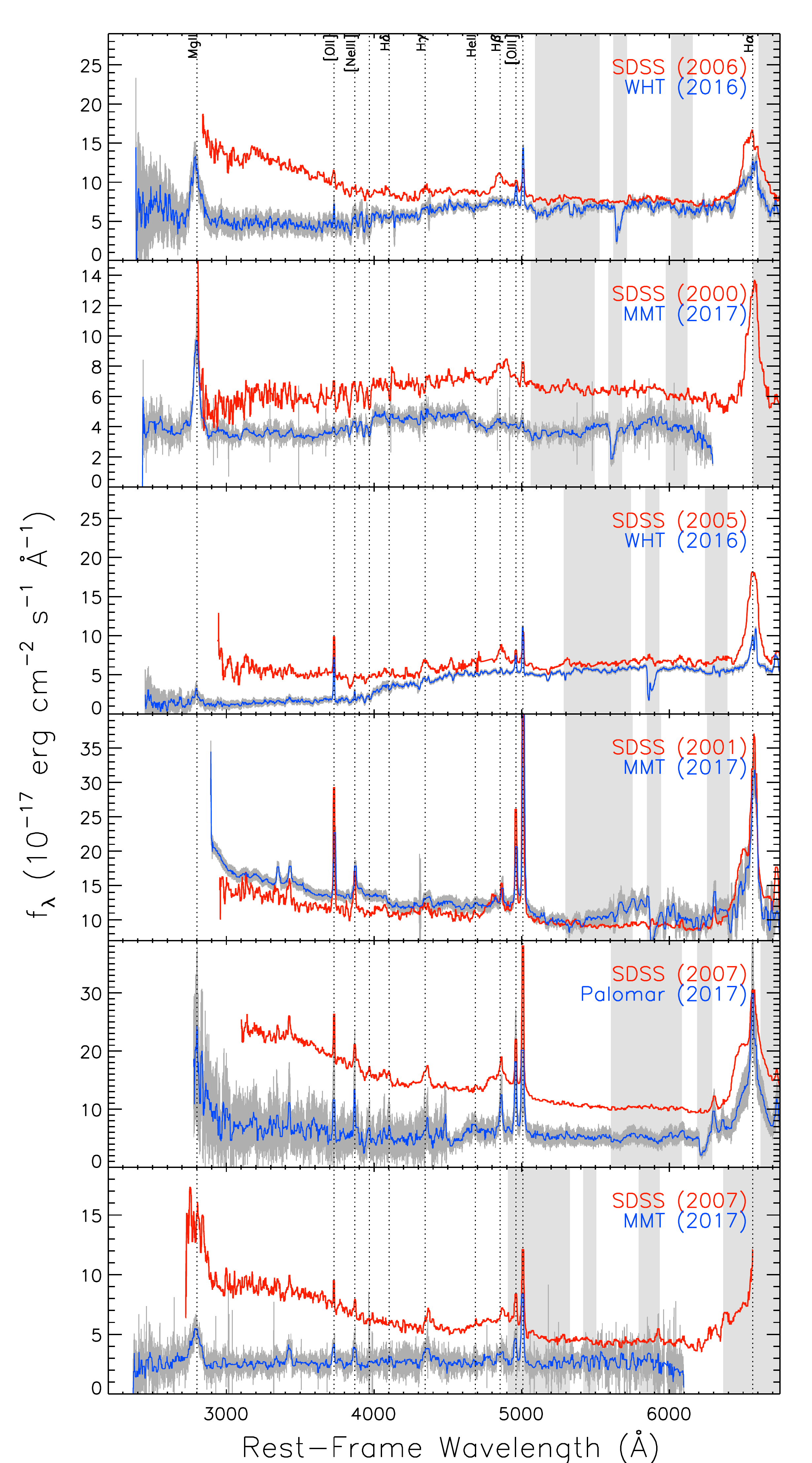}}}
\end{figure}
\begin{figure}[h!]                 
  \ContinuedFloat
  \subfloat[][Fig. A1$c$]{\centerline{
      \includegraphics[scale=.4]{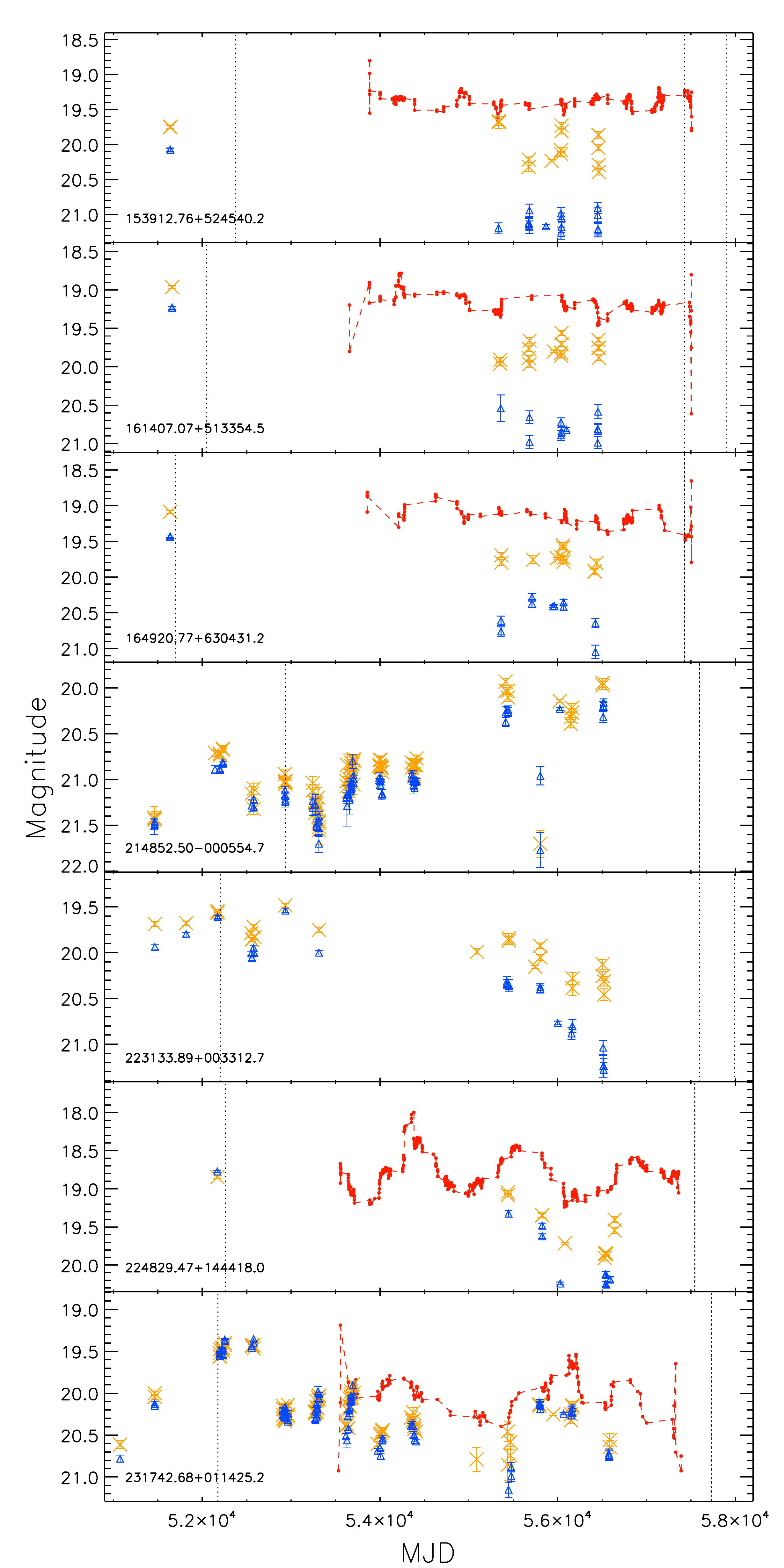}
      \includegraphics[scale=.4]{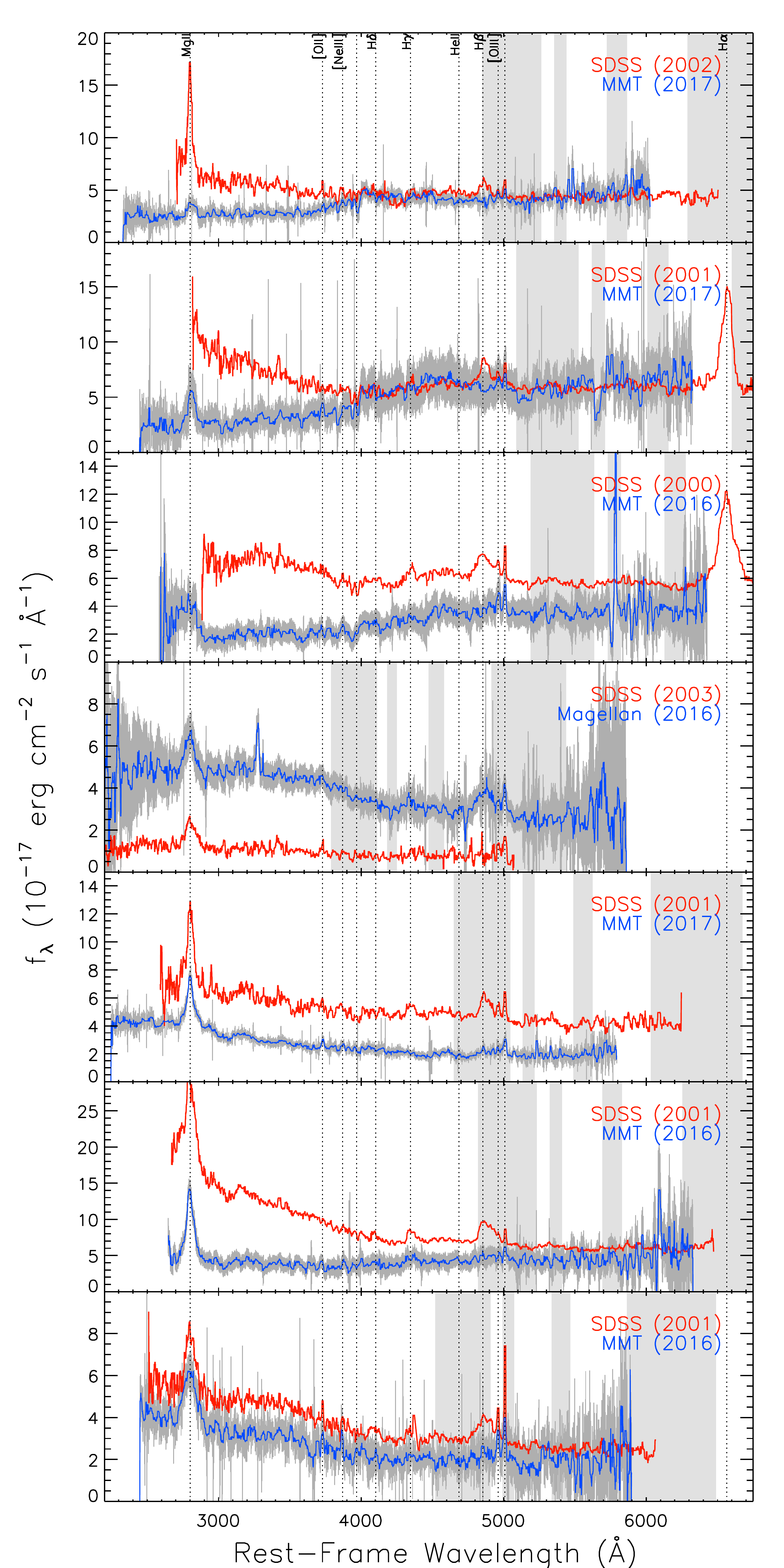}}}
\end{figure}

\setcounter{figure}{1}

\begin{figure}[h!]
\centerline{
\includegraphics[scale=.4]{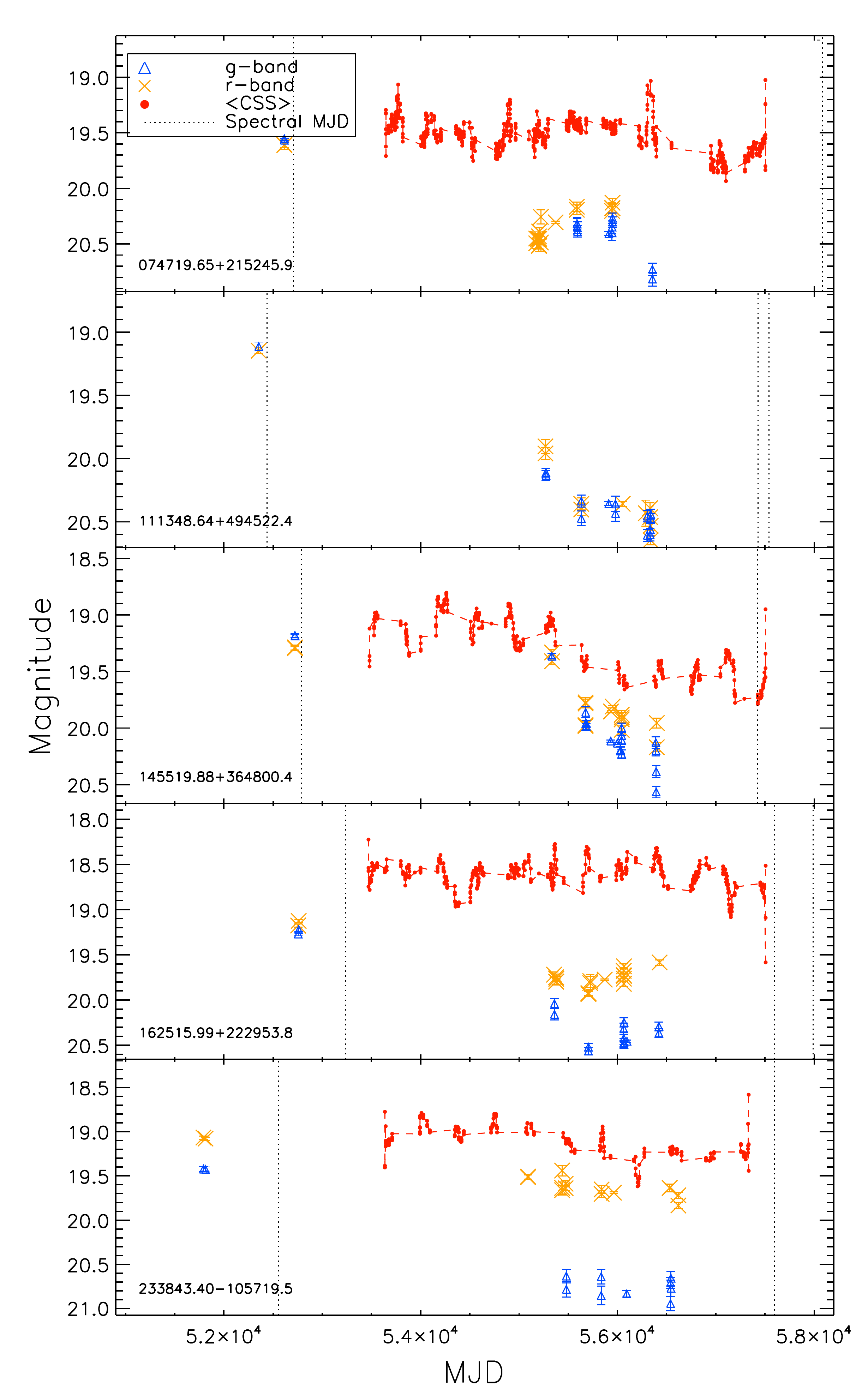}
\includegraphics[scale=.4]{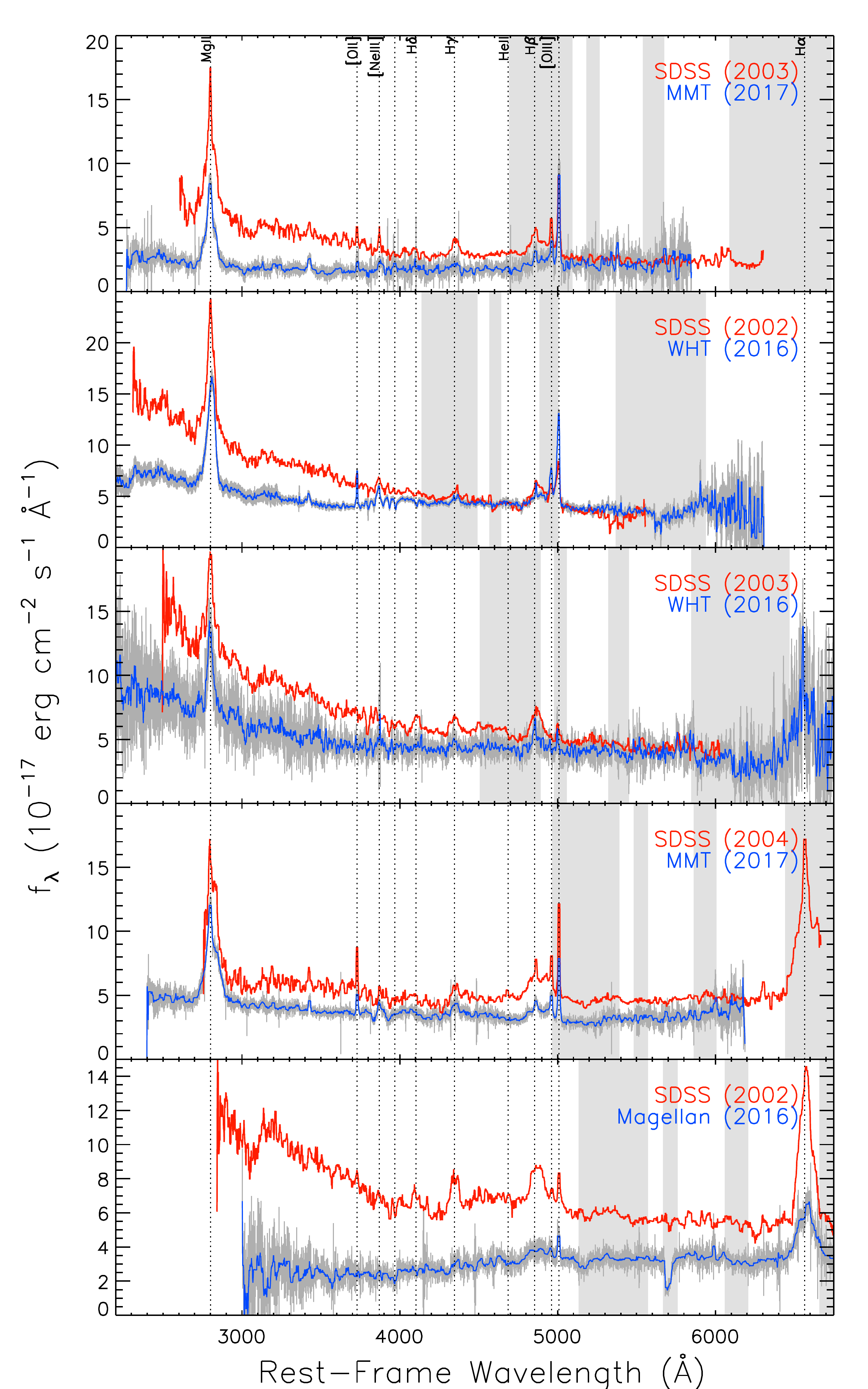}}
\caption{\footnotesize{
Near-CLQs from WHT/MMT/Magellan.  See caption for Figure~\ref{fig:clq1}.
}\label{fig:nearclqs}}
\end{figure}

\section{Previously missed archival, serendipitous CLQ}
\begin{figure}[h]
\centerline{
\includegraphics[scale=.3,angle=90]{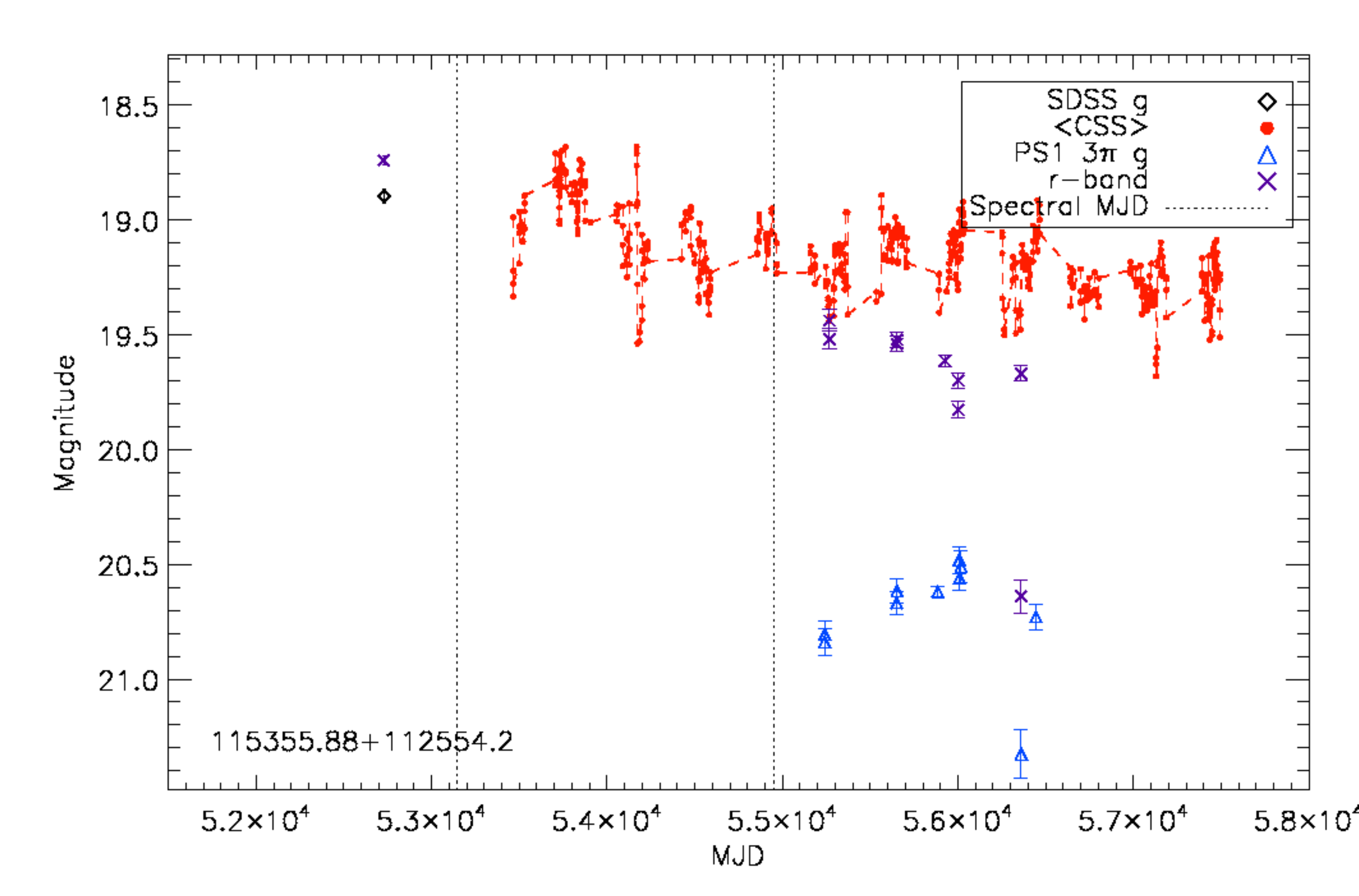}
\includegraphics[scale=.27]{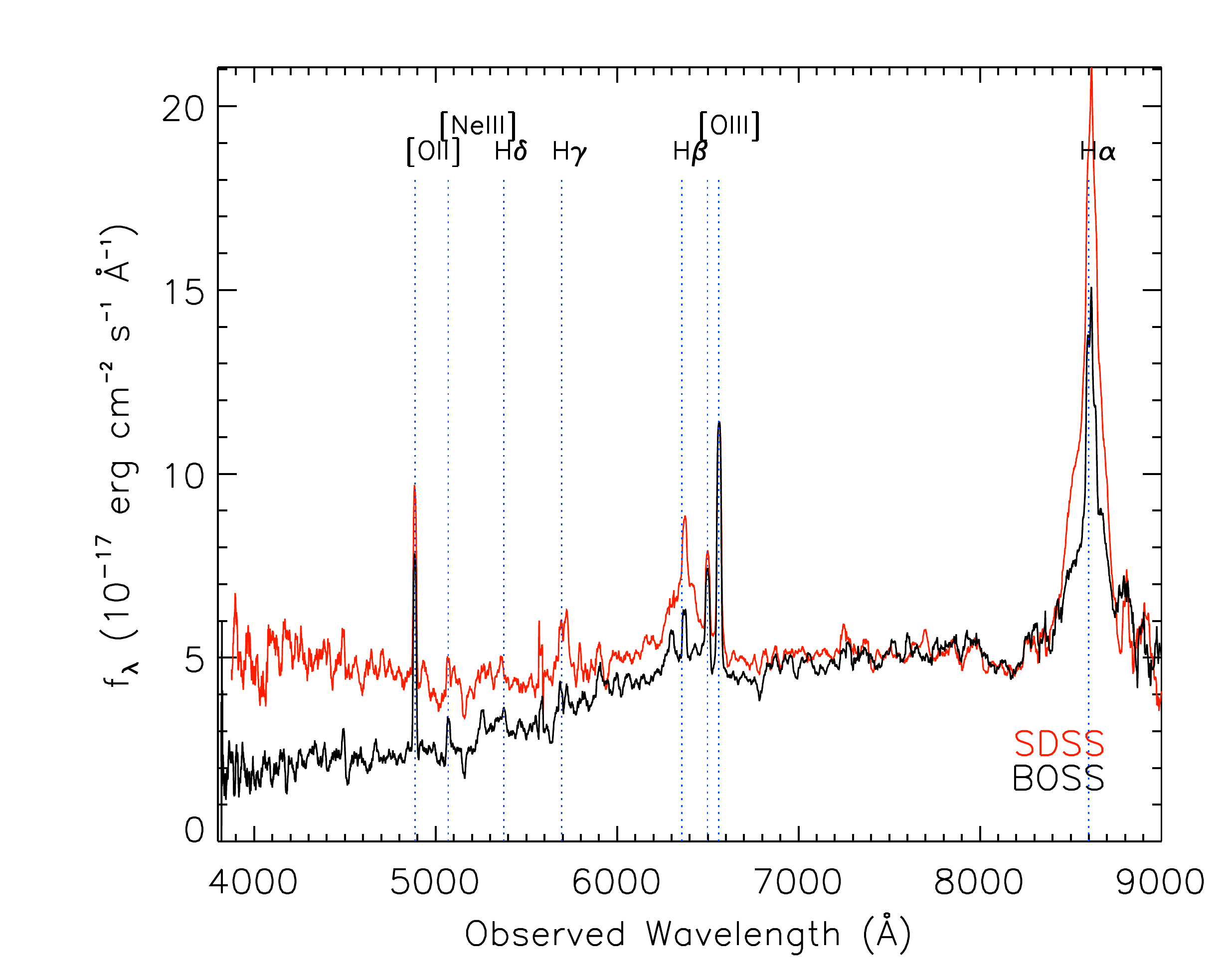}}
\caption{\footnotesize{
\emph{Left:} Light curve for a serendipitous CLQ J115355.88+112554.2
at $z=0.31$ missed by the
\cite{mac15} search.  The existing spectroscopic epochs are indicated by the vertical lines.
\emph{Right:} Existing SDSS and BOSS spectra. The BOSS flux has been
scaled so that the [\ion{O}{3}]$\lambda$5007\AA\ flux matches that in the SDSS spectrum.
}\label{fig:css}}
\end{figure}

In \citet{mac15}, we performed a systematic search for CLQs among quasars with repeat archival spectra in SDSS/BOSS. There,
we only inspected objects that had both SDSS and BOSS epochs, unless the object was in S82; then we inspected all epochs even if they were all in earlier SDSS (pre-BOSS).
One  source, J115355.88+112554.2, had two early SDSS spectral epochs
that already confirm the object as a CLQ (see Figure~\ref{fig:css}).
Although this object was missed by the \cite{mac15} selection since it lacks a
BOSS epoch and is not in S82, it was selected as a CLQ candidate in
the present work.

\section{Comparison to Published CLQs}

SDSS J141324+530527.0 (SBS 1411+533) at $z = 0.456$ \citep{wang18} is
present in DR7Q, but was not selected as a candidate CLQ here since
it only shows a magnitude change of  $-0.86\pm 0.05$ in $g$ from SDSS
to PS1.  
SDSS J012648.08$-$083948.0 \citep{rua15} would also be missed by this
selection since it only varied by $\Delta g=0.78$~mag, but it was not
in our parent sample since a) it is not in DR7Q, and b) it has a BOSS
spectrum. SDSS J233602.98+001728.7 \citep{rua15} is not present in DR7Q,
nor is the CLQ iPTF 16bco \citep{gez17} or
 the CLQs from \cite{yang17}. SDSS J101152.98+544206.4 \citep{run15},
as well as the CLQs from \citet{mac15}, meet the selection criteria
here but already  had confirming BOSS spectra, so we did not follow
them up (except for the two that turned back off described in
\S\,\ref{sec:flicker}). However, we list the CLQs from \citet{run15} and
\citet{mac15} in
Table~\ref{tab:clqcans} for completeness.
The CLQ WISE J105203.55+151929.5 \citep{ste18} is present in DR7Q and
is selected here as a CLQ candidate, though we did not perform
follow-up photometry for this source. 




\bibliography{refs} 
\bibliographystyle{aasjournal}

\end{document}